\documentclass[preprint,12pt,3p, sort&compress]{elsarticle}

\usepackage{graphics}
\usepackage{graphicx}
\usepackage{epsfig}
\usepackage{multirow}
\usepackage{amssymb}
\usepackage{amsthm}
\usepackage{hyperref}
\usepackage{subcaption}
\usepackage{natbib}
\usepackage{amsmath}
\usepackage{booktabs}
\usepackage{longtable}
\usepackage{stackengine}
\usepackage{xfrac}
\usepackage{tabularx}
\usepackage[dvipsnames]{xcolor,colortbl}

\graphicspath{{Figures/}}

\usepackage{bm} 
\usepackage{units} 

\journal{Journal of Sound and Vibration}

\newenvironment{sbmatrix}[1]
{\def\mysubscript{#1}\mathop\bgroup\begin{bmatrix}}
	{\end{bmatrix}\egroup_{\textstyle\mathstrut\mysubscript}}

\begin{document}

\begin{frontmatter}

\title{\textbf{Widening, Transition and Coalescence of Local Resonance Band Gaps in Multi-resonator Acoustic Metamaterials: From Unit Cells to Finite Chains}}


\author[mysecondaryaddress]{A. Stein}

\author[mysecondaryaddress]{M. Nouh}

\author[mysecondaryaddress]{T. Singh\corref{mycorrespondingauthor}}
\cortext[mycorrespondingauthor]{Corresponding author}
\ead{tsingh@buffalo.edu}

\address[mysecondaryaddress]{Department of Mechanical and Aerospace Engineering, University at Buffalo (SUNY),\\ Buffalo, NY 14260-4400, USA}

\begin{abstract}
Local resonance band gaps in acoustic metamaterials are widely known for their strong attenuation yet narrow frequency span. The latter limits the practical ability to implement subwavelength band gaps for broadband attenuation and has motivated novel metamaterial designs in recent years. In this paper, we investigate the behavior of acoustic metamaterials where unit cells house multiple resonating elements stacked in different configurations, aimed at instigating a wide array of wave propagation profiles that are otherwise unattainable. The dispersion mechanics of the multi-resonator metamaterials are developed using purely analytical expressions which depict and explain the underlying dynamics of such systems both at the unit cell level as well as the frequency response of their finite realizations. The framework reveals the mechanism behind the transition of the lower and upper band gap bounds in metamaterials with parallel resonators resulting in a significant band gap widening. The analysis also illustrates the ability of metamaterials with dual-periodic super cells to exhibit a range of dispersion transitions culminating in collapsing solutions of acoustic and optical bands, enabling a coalescence of local resonance band gaps, vanishing resonances, and a number of intriguing scenarios in between.
\end{abstract}


\end{frontmatter}

\newpage

\section{Introduction \label{introduction}}

Over the past two decades, artificially engineered structures, or metamaterials, which exhibit unconventional mechanical properties have garnered a lot of research interest \cite{chen2018review, yu2018mechanical, cummer2016controlling}. Following a sustained investment of efforts in electromagnetic metamaterials, there has been exponential growth relative to the notion of acoustic metamaterials (AMs) where internal architectures give rise to intriguing wave manipulation, guidance, and dissipation \cite{ma2016acoustic}. To date, the use of AMs has spanned a wide range of applications ranging from acoustic lens design \cite{craster2012acoustic}, cloaking \cite{cloak3}, and programmable metasurfaces \cite{tian2019programmable}, to diodes \cite{liang2009acoustic}, logic-based acoustic switches \cite{bilal2017bistable} and nonreciprocal systems \cite{attarzadeh2020beam}. Analogous to their photonic counterparts, the unique properties of acoustic metamaterials are typically shaped by local resonances which are housed within a set of self-repeating cells. The ability of such resonances to localize and absorb incident energy, as first demonstrated by Liu et al. \cite{liu_sonic}, opens up subwavelength band gaps (i.e., regions of forbidden wave propagation) which tend to be (a) mechanically tunable and (b) independent of unit cell size, rendering AMs widely appealing for vibration and noise control applications. Depending on the application, coupling between an outer (host) structure and resonances can take place via internal lattices \cite{baravelli}, axially-oscillating resonators \cite{AlBabaa_PF2}, flexural beams \cite{aladwani2020mechanics}, multilayered inclusions \cite{krushynska2017multilayered}, among others. In their basic form, one-dimensional (1D) acoustic metamaterials can be modeled as a chain of mass-in-mass mechanical oscillators connected via lumped springs. The simplicity of such a model lends insight into the mechanics of wave dispersion and the resonance hybridization which culminates in the opening of a narrow band gap along the frequency axis. 

Despite the appeal of a very strong, mechanically-tunable band gap that can extend to low frequency regimes, the potential of local resonance band gaps remains largely untapped as a result of their narrow frequency range, unlike wide Bragg scattering band gaps in periodic crystals \cite{hussein_transition, BG_synthesis_meccanica}. As a result, a number of recent efforts have emerged with a common theme of exploring novel designs in search of AM configurations that exhibit broader wave dispersion features. Notable among these are designs which involve interactions of neighboring resonators, either in spring-mass form \cite{hu2017acoustic} or elastic beams with interconnections \cite{beli2018wave}. Other efforts have investigated the concepts of non-local resonances \cite{depauw2018metadamping}, dual chain coupling \cite{hajarolasvadi2021dispersion}, electrically-resonant systems \cite{callanan2021uncovering}, origami designs \cite{pratapa2018bloch}, as well as inertial amplification mechanisms \cite{inerters_Singh_JAP}, as ways to achieve versatile acoustic metamaterials. Aside from Bloch- and Floquet-based methods which project the behavior of an infinite (hypothetical) metamaterial from a single cell, there has also been a growing interest in the dynamics of finite AMs and the different ways in which dispersion mechanics evolve and manifest themselves in the response of finite systems. This has led to the use of tools such as confinement effects theory \cite{ren2007theory}, modal analysis \cite{sugino2016mechanism}, the phase-closure principle \cite{hvatov2015free}, and layer-multiple-scattering techniques \cite{sainidou2008surface} to further understand the underlying behavior of finite dispersive structures. Likewise, continuous-fraction block diagram models of finite periodic structures have provided an interpretation of band gap formation mechanisms in phononic \cite{NouhSinghJASA}, permutative, \cite{NouhSinghRSPA}, as well as locally resonant metamaterials \cite{AlBabaa.2017, NouhSinghSPIE}.

\newpage

The pursuit of new resonant mechanisms has demonstrated that local resonance band gaps do not necessarily conform to narrow frequency ranges. Of particular interest to this work is the notion of using multi-resonator unit cells to instigate new characteristics. Most recently, resonators connected in parallel \cite{tian2019multi, zhao2021multi} or in series \cite{aladwani2021strategic} have been shown to exhibit enhanced band gaps in addition to improved dissipation properties when analyzed in their dissipative form. In tandem, resonators assembled in a \textit{super cell} dual-periodic configuration have demonstrated the possibility of resonance coupling between two distinct tuning frequencies as shown in the work of Gao and Wang \cite{gao2020ultrawide}. Although promising ideas, most efforts thus far have focused on exploiting these designs for practical applications (e.g., double negativity \cite{li2021wave}, noise control \cite{roca2021multiresonant}, and broadband sound transmission loss \cite{roca_hussein_dual2021}) which, while of profound importance, do not provide a comprehensive explanation of the underlying physics behind multi-resonator configurations or the role they play in the onset of unprecedented features in the dispersion diagrams thereof. As a result, the following fundamental questions remain unanswered: (1) Why do some design variations give rise to enhanced band gap properties while others don't? (2) What are the different elastic wave dispersion profiles that can be triggered using such designs? and (3) How do phenomena such as band gap widening, merging of two local resonance band gaps, emergence, shifting, or disappearance of Bragg band gaps form in a finite metamaterial in order to eventually comply with the unit cell prediction? The primary aim of this work is to answer all of these questions through a two-part framework beginning with traditional dispersion diagrams, providing a mathematical interpretation of a wide array of dispersion scenarios associated with various multi-resonator metamaterials, and ending with a frequency-domain closed-form transfer function approach which connects these scenarios to their finite counterparts.

The paper is organized as follows: The first section focuses on unit cell analysis of two main variations of multi-resonator lumped acoustic metamaterials. The first includes $r>1$ resonators connected in parallel while the second is comprised of a super cell in the form of two mass-in-mass cells housing distinct resonances and connected in series. Generalized expressions are derived for the dispersion relations, band gap bounds, and other metrics of both systems which are then investigated numerically using a set of defined parameters. An interpretation of the different dispersion profiles is provided which addresses the necessary conditions for band gap widening as well as the distribution of local resonance modes and corresponding flat branches in unit cells with parallel resonators. Following which, the super cell configuration is extensively analyzed and a discussion of the different types and scenarios of local resonance band gap merging (in addition to the corresponding appearance, shifting, and vanishing of Bragg band gaps) is presented. In the second section, closed-form finite models of the same multi-resonator metamaterials are derived based on a structural dynamics model followed by a detailed transfer function approach. Finally, we validate the infinite chain predictions graphically using a set of frequency response functions. The presented work provides a pathway for local-resonance-based acoustic metamaterials to play a more robust role in applications requiring broadband attenuation of sound and vibration loads, particularly using recent advances in physics-based inverse design and multi-objective optimization algorithms.

\newpage

\section{Multi-resonator Acoustic Metamaterials: \textit{Unit Cell Analysis}}
\label{section:Dispersion}

\subsection{Parallel identical resonators}
\label{subsection:Dispersion_Analysis_Parallel}

\begin{figure}[h!]
	\centering
	\includegraphics[width=\textwidth]{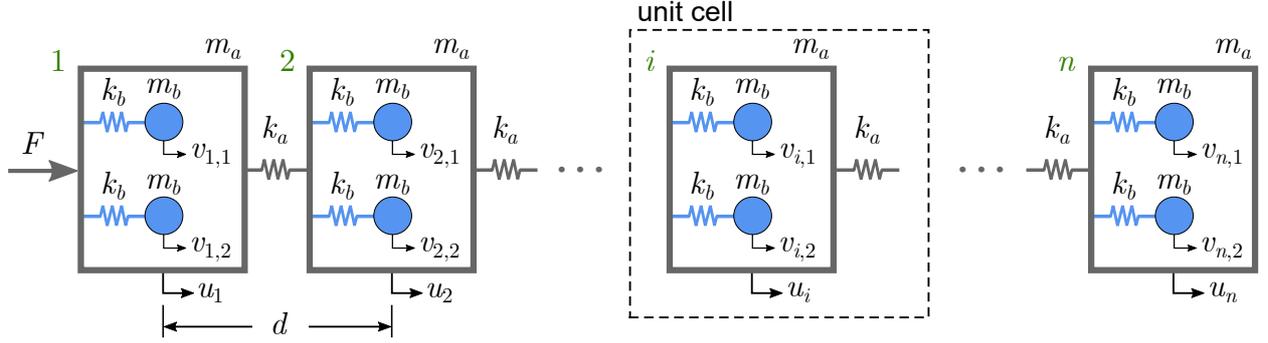}
	\caption{Lumped parameter model of a multi-resonator acoustic metamaterial (AM). The diagram shows a system of $n$ unit cells and $r=2$ identical resonators per cell, connected in parallel and housed inside an outer mass $m_a$. Each resonator consists of a mass $m_b$ and stiffness $k_b$, the outer masses of the chain are connected via identical springs $k_a$ and are a distance $d$ apart.}
 \label{fig:parallel_lumped_mass_system}
\end{figure}

As mentioned earlier, in their simplest form, 1D acoustic metamaterials (AMs) can be modeled as a series of lumped masses with locally resonating elements (see Fig.~\ref{fig:parallel_lumped_mass_system}). For an AM with two identical internal resonators per unit cell (in a parallel arrangement), the dynamics of the $i^{\text{th}}$ cell can be represented as follows:
\begin{subequations}
\begin{align}
m_a\ddot{u}_i + (2k_a+2k_b)&u_i - k_bv_{i,1} - k_bv_{i,2} - k_au_{i-1} - k_au_{i+1} = 0\\
&m_b\ddot{v}_{i,1} + k_bv_{i,1} - k_bu_i = 0\\
&m_b\ddot{v}_{i,2} + k_bv_{i,2} - k_bu_i = 0
\end{align}
\label{eq:AM_2r}
\end{subequations}
where $u_i$, $v_{i,1}$, and $v_{i,2}$ represent the displacements of the outer mass, first resonator, and second resonator, respectively. The double dots denote an acceleration term, $m_a$ and $m_b$ represent the outer and resonator masses, respectively, $k_b$ represents the resonator spring, and $k_a$ is the stiffness coefficient of the spring connecting two adjacent unit cells. Employing the Bloch-wave solution \cite{farzbodanalysis}, the following form can be derived from Eq.~(\ref{eq:AM_2r}):
\begin{subequations}
\begin{align}
(-\Omega^2\omega_b^2m_a +& 2k_a[1-\cos \tilde{k}]+2k_b) \tilde{u}- k_b\tilde{v}_1  - k_b\tilde{v}_2  = 0 \\
&-\Omega^2\omega_b^2m_b\tilde{v}_1 + k_b\tilde{v}_1 - k_b\tilde{u} = 0\\
&-\Omega^2\omega_b^2m_b\tilde{v}_2 + k_b\tilde{v}_2 - k_b\tilde{u} = 0
\end{align}
\label{eq:Bloch_2r}
\end{subequations}
where $\tilde{u}$, $\tilde{v}_1$, and $\tilde{v}_2$ represent the respective periodic displacement fields, $\tilde{k}=kd$ is the dimensionless wavenumber (product of the cell spacing $d$ and the angular wavenumber $k$), $\Omega =\frac{\omega}{\omega_b}$ is the dimensionless frequency, where $\omega$ is the wave frequency, and $\omega_b=\sqrt{k_b/m_b}$ is the stand-alone natural frequency of an individual resonator. A generalization of Eq.~(\ref{eq:Bloch_2r}) for a multi-resonator metamaterial with $r$ identical resonators produces:
\begin{equation}
\begin{bmatrix}
-\Omega^2\omega_b^2 m_a  + 2 k_a[1-\cos \tilde{k}] + r k_b  & -k_b & \dots & -k_b \\
-k_b & k_b - \Omega^2 \omega_b^2 m_b \\
\vdots& & \ddots\\
-k_b & & & k_b - \Omega^2 \omega_b^2 m_b
\end{bmatrix}
\begin{bmatrix}
\tilde{u}\\
\tilde{v}_1\\
\vdots\\
\tilde{v}_r
\end{bmatrix}
=
\mathbf{0}.
\label{eq:disp_lumped}
\end{equation}

\begin{figure}[h!]
\centering
	\includegraphics[width=0.94\textwidth]{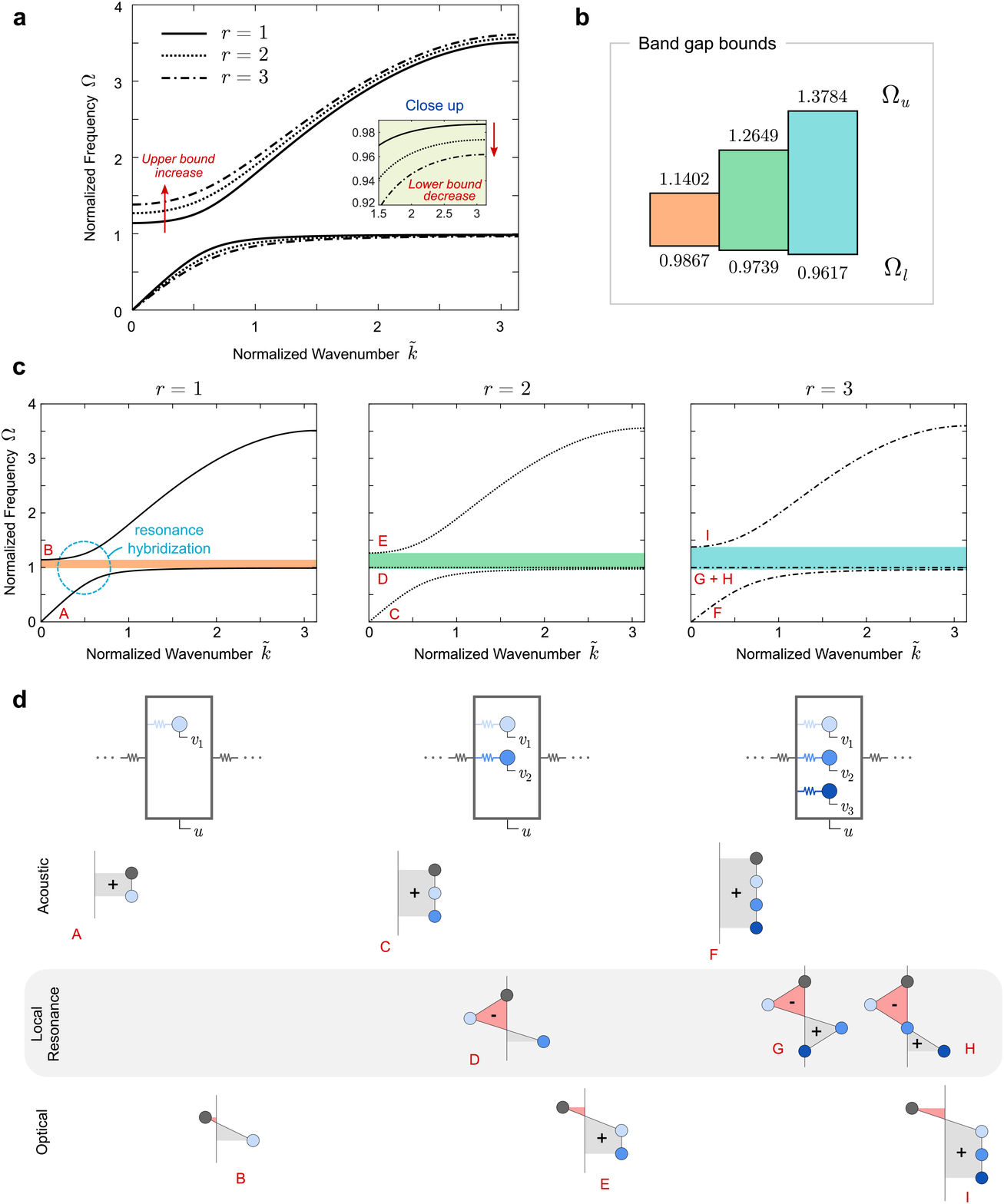}
	\caption{(a) Dispersion diagram for a lumped acoustic metamaterial with $r=1$, $2$, and $3$ identical resonators in parallel. The parameters used are as follows: $m_a=1$, $m_b=0.3$, $k_a=4.8e9$, and $k_b=0.1k_a$. Close up inset shows lower band gap bound. (b) Upper and lower band gap bounds, $\Omega_u$ and $\Omega_l$, corresponding to the three cases. (c) Individual representation of the three dispersion diagrams. Shaded areas indicated band gap regions. (d) Acoustic, optical, and local resonance unit cell mode shapes for each case.} 
	\label{fig:dispersion_curves}
\end{figure}

Non-trivial solutions of Eq.~(\ref{eq:disp_lumped}) can be obtained by computing the determinant of the left-hand side matrix for a given set of parameters. The latter yields a polynomial of degree $2(r+1)$ which can be used to obtain the AM's dispersion relation, i.e., $\Omega(\tilde{k})$. Using the following parameters: $m_a=1$, $m_b=0.3$, $k_a=4.8e9$, and $k_b=0.1k_a$, Fig.~\ref{fig:dispersion_curves} shows the AM's dispersion curves obtained using such approach for $r=1$, $2$, and $3$, respectively. Fig.~\ref{fig:dispersion_curves}a reveals an interesting trend where the AM's local resonance band gap expands from both ends as the number of identical parallel resonators increases. The upper bound of the band gap moves to a higher frequency in a clear and significant manner, while the lower bound shifts to a slightly lower frequency as seen in the close up inset. A numerical quantification of the band gap bounds in the three cases is displayed in Fig.~\ref{fig:dispersion_curves}b. This behavior will be explained analytically in the forthcoming sections as well as validated in the finite response of the multi-resonator metamaterials in section~\ref{subsection:BG_widening_finite}. For now, we focus on Fig.~\ref{fig:dispersion_curves}c where the left panel depicts the well-known dispersion diagram of a single-resonator lumped acoustic metamaterial (also known as the mass-in-mass system). A hybridization between the resonant inclusion and the bulk structure culminates in an avoided-crossing between the bulk dispersion branch and the (flat) resonance, and opens up a narrow subwavelength band gap; as first reported by Liu et al.~\cite{liu_sonic}. With the addition of a second resonator ($r=2$), in addition to the band gap expansion, the determinant obtained from Eq.~(\ref{eq:disp_lumped}) now yields a polynomial of degree $2(r+1) = 6$, resulting in three dispersion branches (and three negative ones which are rejected). The three branches, shown in the middle panel of Fig.~\ref{fig:dispersion_curves}c, represent the unit cell's acoustic mode, optical mode, as well as a resonance mode at $\Omega=1$. For $r=3$, these become two coinciding resonance modes and so forth. To generalize, a cell comprised of $r$ identical parallel resonators will have $(r-1)$ flat resonance branches. The unit cell (or Bloch-wave) modes are obtained by casting the dispersion relation as an eigenvalue problem and solving for the eigenvectors, $\mathbf{q}$, or alternatively by computing the null space of the left hand side matrix in Eq.~(\ref{eq:disp_lumped}). The mode shapes are visually illustrated in Fig.~\ref{fig:dispersion_curves}d following the format $\mathbf{q} = [\tilde{u} \:\: \tilde{v}_1 \:\:... \:\: \tilde{v}_r]^T$. It can be seen that, irrespective of the number of resonators $r$, the host (outer) mass only oscillates within the acoustic and optical modes. In between, local resonance modes (the number of which depends on the value of $r$) all capture different relative motions between the individual resonators with $\mathbf{q}(1) = 0$ being a common feature in all of them. A pattern emerges which can be used to generalize this behavior to a multi-resonator AM of any value of $r>1$. We define $\mathbf{q}_a$ as an $(r+1) \times 1$ column vector representing the AM's acoustic mode, $\mathbf{q}_o$ as an $(r+1) \times 1$ column vector representing the AM's optical mode, and $\mathbf{q}_r$ as an $(r+1) \times (r-1)$ matrix depicting the different possible local resonance modes, such that for any multi-resonator lumped AM, the Bloch-wave mode matrix $\mathbf{Q} = [\mathbf{q}_a \:\: \mathbf{q}_r \:\: \mathbf{q}_o]$ takes the form:
\begin{equation}
\mathbf{Q} = 
	\begin{sbmatrix}{I_{(r-1) \times (r-1)}}
\overbrace{1}^\text{acoustic} & 0  & \cdots &  0 & \overbrace{-e}^\text{optical}\\
1 & -1 & \cdots & -1 & 1\\
\cline{2-4}
\vdots & \multicolumn{1}{|c}{1} & & \multicolumn{1}{c|}{} & \vdots\\
\vdots & \multicolumn{1}{|c}{} & \ddots &  \multicolumn{1}{c|}{}& \vdots\\
1 & \multicolumn{1}{|c}{} & &  \multicolumn{1}{c|}{1} & 1\\
\cline{2-4}
	\end{sbmatrix}
\label{eq:disp_lumped_modes}
\end{equation}
where $e>0$ and changes depending on the number of resonators. For the given parameters, $e=0.6$ and $0.9$ corresponding to $r=2$ and $3$, respectively. 

As will be shown later, the presence of these local resonance modes (flat branches) does not affect band gap formation in finite lumped acoustic metamaterials. The band gap width remains dictated by the acoustic and optical modes and, as such, its bounds can be calculated accordingly. Consequently, using the system of equations in Eq.~(\ref{eq:disp_lumped}) and implementing a row reduction, the following reduced dispersion relation can be obtained for unit cells with \textit{identical} parallel resonators: 
\begin{equation}
(-\Omega^2\omega_b^2 m_b + k_b)(-\Omega^2\omega_b^2 m_a + 2k_a[1-\cos \tilde{k}]+rk_b) - rk_b^2 = 0
\label{eq:dispersion_analysis_general}
\end{equation}

\subsubsection{Local resonance band gap widening}
\label{subsection:Dispersion_Analysis_widening}

A commonly used approximation is that the local resonance band gap in a lumped acoustic metamaterial starts at $\Omega=1$, i.e., at the natural frequency of the internal absorber \cite{Pai.2014, Peng.2015}. However, a close inspection of Eq.~(\ref{eq:dispersion_analysis_general}) reveals that the actual lower bound of the band gap can deviate from this value and start at lower frequencies depending on the chosen parameter values (See Fig.~3 in Ref.~\cite{AlBabaa.2017} for a detailed illustration of such deviation). Owing to the shape of an AM's dispersion diagram, the lower bound of the band gap of the multi-resonator metamaterial can be analytically determined by setting $\tilde{k}=\pi$ in Eq.~(\ref{eq:dispersion_analysis_general}). Neglecting negative solutions and taking the smaller of the two positive ones, we get:
\begin{equation}
\Omega_{l} = \left( \frac{\left(4k_am_b + rk_bm_b + k_bm_a\right)}{2\omega_b^2m_am_b}  -\sqrt{\frac{\left(4k_am_b + rk_bm_b + k_bm_a\right)^2-16k_ak_bm_am_b}{4(\omega_b^2m_am_b)^2}}\right)^{1/2}
\label{eq:Omega_l_dispersion_analyis_unsimplified}
\end{equation}
Setting $m_r=m_b/m_a$, $k_r=k_b/k_a$ and $\Gamma = m_r/k_r$, Eq.~(\ref{eq:Omega_l_dispersion_analyis_unsimplified}) simplifies to:
\begin{equation}
\Omega_{l} = \frac{1}{\sqrt{2}} \sqrt{\left(4\Gamma+ rm_r + 1\right) - \sqrt{\left(4\Gamma+ rm_r + 1\right)^2-16\Gamma}}
\label{eq:Omega_l_dispersion_analyis_closed_form}
\end{equation}

On the other hand, using Eq.~(\ref{eq:dispersion_analysis_general}) and setting $\tilde{k}=0$ yields the band gap's upper bound $\Omega_u$, which is given by:
\begin{equation}
\Omega_{u} = \sqrt{r m_r + 1}
\label{eq:Omega_u_dispersion_analyis_closed_form}
\end{equation}

The effect of the number of resonators $r$ is evident in Figs.~\ref{fig:dispersion_curves}a and b. It can be seen that $\Omega_u$ increases while $\Omega_l$ decreases with an increase in $r$, resulting in a widening of the local resonance band gap from both ends as can be seen in Fig.~\ref{fig:dispersion_curves}a and the provided close-up inset. Specifically, the upper bound of the band gap solely depends on the number of resonators and the mass ratio between a single resonator and the outer mass of a multi-resonator metamaterial. The increase in the upper bound of the band gap with an increasing $r$ is a straightforward consequence of the growth of the term $r m_r$ in Eq.~(\ref{eq:Omega_u_dispersion_analyis_closed_form}). On the other hand, the downshift of $\Omega_l$ with an increasing $r$ is a bit more complicated and depends on both the mass and stiffness ratios in addition to $r$, as evident by Eq.~(\ref{eq:Omega_l_dispersion_analyis_closed_form}).

\subsubsection{Lack of band gap widening}
\label{subsection:Dispersion_Analysis_no_widening}

\begin{figure}[h!]
\centering
	\includegraphics[width=\textwidth]{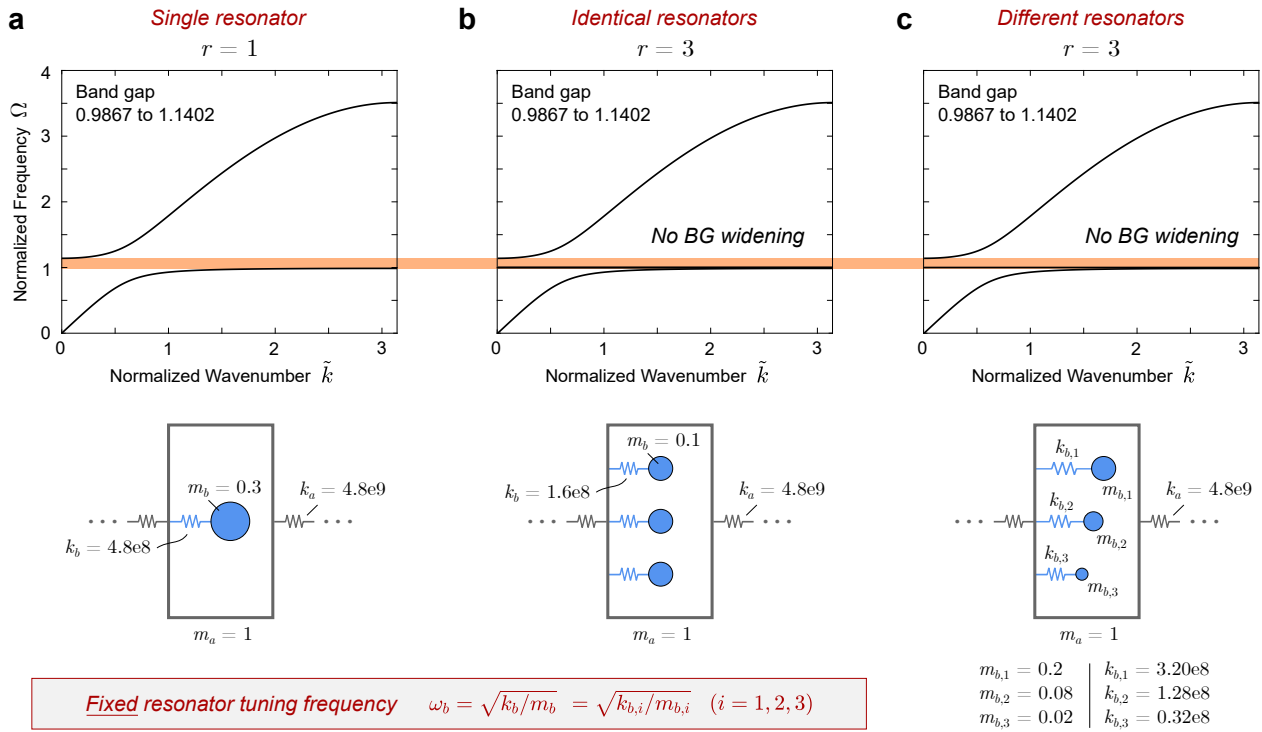}
	\caption{Dispersion diagram for a lumped acoustic metamaterial with a constant total resonator mass of $r m_b = 0.3$. (a) Single resonator: $r=1$ and $m_b = 0.3$, (b) Three identical resonators: $r=3$ each of mass $m_b = 0.1$, and (c) Three different resonators: $r=3$ having the following masses $m_{b,1} = 0.2$, $m_{b,2} = 0.08$, and $m_{b,3} = 0.02$. In all three systems, resonator stiffnesses are chosen such that the tuning frequency $\omega_b = \sqrt{k_b/m_b}$ or $\sqrt{k_{b,i}/m_{b,i}}$ (where $i=1,2,3$) remains unchanged.}
	\label{fig:dispersion_analysis_constant_mass}
\end{figure}

Figure~\ref{fig:dispersion_analysis_constant_mass} shows the dispersion diagram changes resulting from using multiple parallel resonators while keeping the total resonator mass per cell unchanged (i.e., $r m_b = 0.3$), while adjusting the resonator stiffnesses to maintain a constant tuning frequency (i.e., $\omega_b = \text{constant}$). In Fig.~\ref{fig:dispersion_analysis_constant_mass}a, the entire resonator mass is contained within a single resonator with $r=1$ and $m_b = 0.3$. The rest of the parameters are indicated on the schematic diagram in the lower panel (and are kept similar to those used in Figure~\ref{fig:dispersion_curves} for convenience). This represents the conventional mass-in-mass system which is provided as a benchmark for comparison purposes. In Fig.~\ref{fig:dispersion_analysis_constant_mass}b, the resonator mass is distributed evenly among 3 resonators, while in Fig.~\ref{fig:dispersion_analysis_constant_mass}c the resonator mass is distributed unevenly among 3 resonators. As seen in the figures, the band gap starting and ending frequencies, and consequently width, are the same for all three systems. The presence of identical resonators (Fig.~\ref{fig:dispersion_analysis_constant_mass}b) facilitates the simplification of the motion equations of the multi-resonator cell and makes it easier to explain this behavior using the expressions derived earlier. Here, $\Omega_u$ does not change since any multiple of $r$ in Eq.~(\ref{eq:Omega_u_dispersion_analyis_closed_form}) is cancelled out as a result of dividing $m_b$ by the same multiple. The ratio $\Gamma$ also remains unchanged in Eq.~(\ref{eq:Omega_l_dispersion_analyis_closed_form}) for the same reason. 

As a result of this, we note that the band gap widening phenomenon depicted in section~\ref{subsection:Dispersion_Analysis_widening} requires an increase of resonator mass, either by directly increasing the mass of a single resonator or by stacking multiple identical resonators in parallel as has been shown here. While the former approach provides the widening effect, it assumes that the user has complete freedom to choose any resonator mass. It also assumes that the user has access to any number of resonators when assembling the finite metamaterial. However, as will be shown later in section~\ref{subsection:BG_widening_finite} and Fig.~\ref{fig:BG_comparision_n12_r1_n4_r3}, we make the case that given a limited number of resonators with a specific mass, we are better off stacking these resonators in a fewer number of cells by exploiting the parallel configuration. This arrangement proves to be superior both from the standpoint of band gap width as well as the number of pass-band resonant peaks exhibited over the same frequency range. Furthermore, despite the increase in unit cell mass, it actually reduces the ``overall" mass of the finite metamaterial due to the significant reduction in the chain length and the need for much fewer outer masses.

\subsubsection{Non-identical parallel resonators with multiple band gaps}

\begin{figure}[h!]
\centering
	\includegraphics[width=\textwidth]{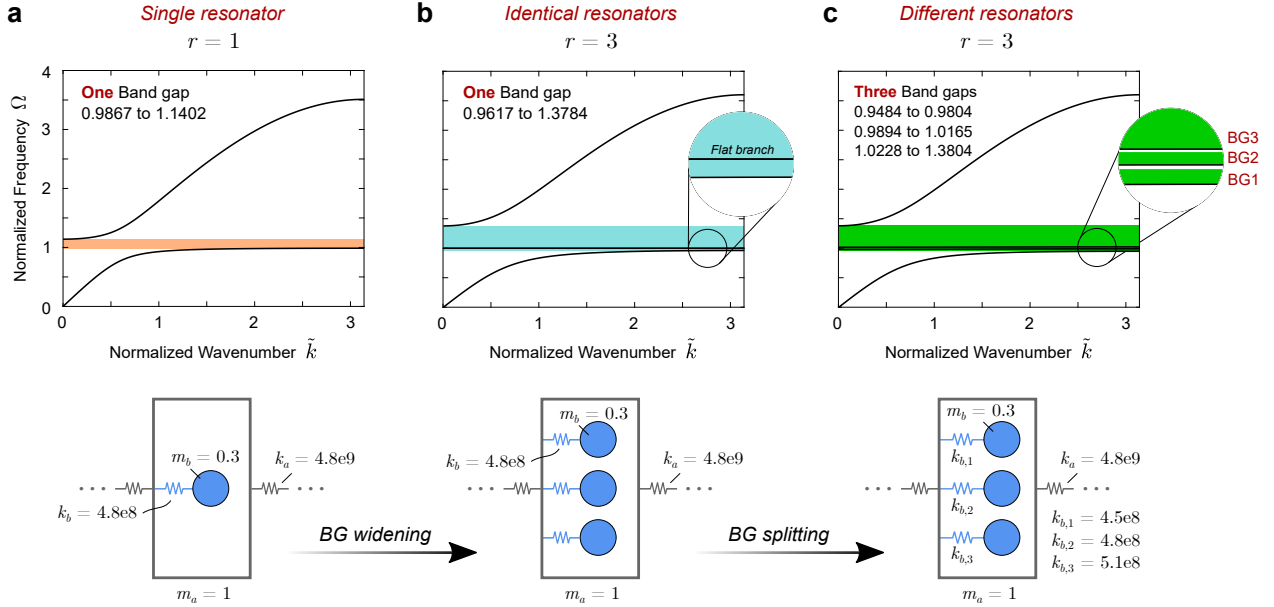}
	\caption{Dispersion diagram for a lumped acoustic metamaterial with (a) Single resonator: $r=1$ with $m_b = 0.3$ and $k_b = 4.8e8$, (b) Three identical resonators: $r=3$ each of mass $m_b = 0.3$ and $k_b = 4.8e8$, and (c) Three different resonators: $r=3$ each of mass $m_b = 0.3$ and having the following stiffnesses $k_{b,1} = 4.5e8$, $k_{b,2} = 4.8e8$, and $k_{b,3} = 5.1e8$. The system shown in (b) reveals a widening of the band gap incurred by the single resonator system, while the one in (c) shows splitting into three distinct band gaps spanning the frequency ranges listed on the figure.}
	\label{fig:dispersion_analysis_increased_mass}
\end{figure}

To conclude the analysis of unit cells comprising resonators arranged in parallel, we briefly present the effect of including non-identical resonators. Unlike the case shown in Fig.~\ref{fig:dispersion_analysis_constant_mass}c, we remove the mass constraint in addition to allowing the resonator stiffnesses to vary; therefore generating a set of resonators with completely different tuning frequencies. Figure~\ref{fig:dispersion_analysis_increased_mass} displays the dispersion behavior of an $r=3$ unit cell with three distinct resonator stiffnesses, $k_{b,i}$ where $i = 1$, $2$, and $3$. As a result, this configuration yields three distinct tuning frequencies, at $\omega_{b,i} = \sqrt{k_{b,i}/m_b}$, which correspond to three different band gaps indicated by the shaded regions and the frequency ranges listed on Fig.~\ref{fig:dispersion_analysis_increased_mass}c. Despite the narrow frequency ranges separating the three band gaps, it should be emphasized that this configuration provides separate back-to-back band gaps. In other words, the dispersion branches separating the three band gaps in this scenario are not flat as can be inferred from the starting and ending frequencies of each of the three gaps; unlike the case of the identical resonators which is shown for comparison in Fig.~\ref{fig:dispersion_analysis_increased_mass}b. The dispersion diagrams of both cases otherwise look largely similar. Finally, this behavior (i.e., BG \textit{widening} vs. \textit{splitting}) also manifests itself in the finite realizations of the unit cells shown in Figs.~\ref{fig:dispersion_analysis_increased_mass}b and c, respectively. In the former, the frequency response is expected to show a single uninterrupted attenuation region while the latter will reveal three distinct band gaps with resonant peaks in between representing the finite vibrational modes which span the narrow pass bands in between the three gaps.

\subsection{The super cell configuration}
\label{subsection:Dispersion_Analysis_Supercell_Approach}

\begin{figure}[h!]
\centering
	\includegraphics[width=\textwidth]{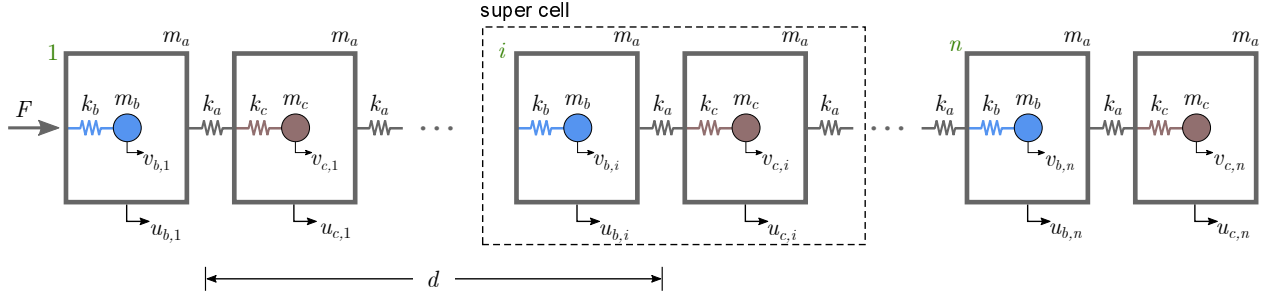}
	\caption{Lumped parameter model of a multi-resonator acoustic metamaterial with a dual periodic super cell design. The diagram shows a system of $n$ super cells consisting of two identical outer masses $m_a$ connected in series via a spring of stiffness $k_a$, and two resonators $m_b$ and $m_c$, which are connected to their respective outer masses via the stiffnesses $k_b$ and $k_c$. $d$ is the distance between two consecutive super cells.}
	\label{fig:Supercell_Sketch_Lumped_masses}
\end{figure}

A super cell of an acoustic metamaterial typically refers to a larger unit cell which is comprised of a number of smaller building blocks, or subcells, and self-repeats along one or more spatial dimensions. Although the super cell can be perceived as a non-conventional unit cell which is often used to give rise to novel features, the constitutive subcells themselves may take conventional forms reminiscent of traditional phononic crystals or locally resonant metamaterials. Given the focus of this work on multi-resonator AMs, we provide here a comprehensive analysis of a class of super cells which connects two (generally) different locally resonant cells which are connected in series. A schematic diagram of the lumped parameter model of the system is shown in Fig.~\ref{fig:Supercell_Sketch_Lumped_masses}. This dual periodic super cell configuration was recently introduced by Gao and Wang as a mechanism which provides an ultrawide low-frequency band gap based on resonance coupling \cite{gao2020ultrawide}. The idea has spurred a number of subsequent efforts which make use of this phenomenon in the domain of noise control and acoustic transmission loss \cite{roca2021multiresonant, roca_hussein_dual2021}. However, as will be shown here, despite being a very promising concept, resonance coupling in the super cell configuration is a nontrivial mathematical problem which requires certain existence conditions. Furthermore, the AM design shown in Fig.~\ref{fig:Supercell_Sketch_Lumped_masses} exhibits a broad range of different (and very intriguing) dispersion characteristics which culminate in various Bragg and local resonance interactions giving rise to coalescence, shifting, and/or vanishing of the different band gaps.

\begin{figure}[h!]
\centering
	\includegraphics[width=0.98\textwidth]{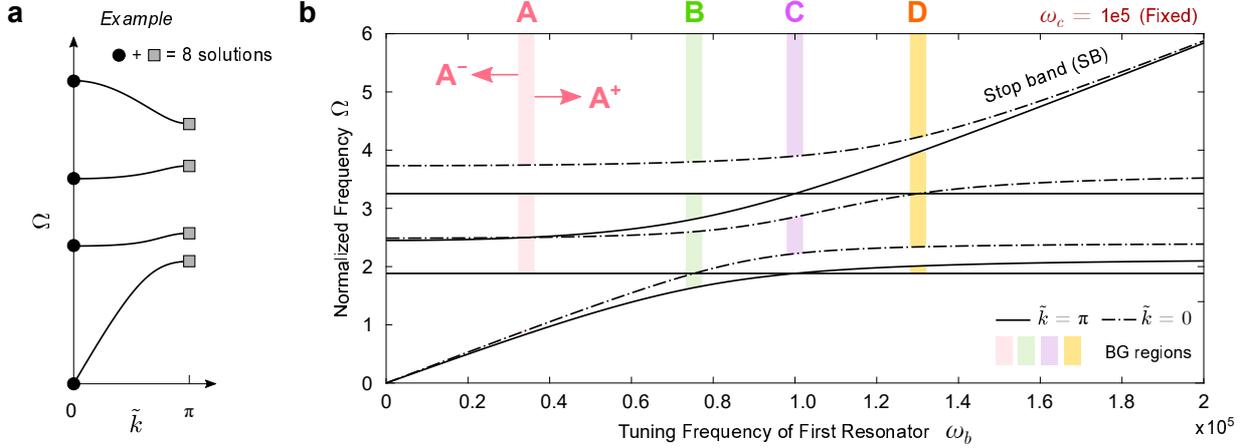}
	\caption{Dispersion scenarios associated with the dual periodic super cell configuration shown in Fig.~\ref{fig:Supercell_Sketch_Lumped_masses}. (a) An example dispersion diagram showing eight possible solutions of Eq.~(\ref{eq:disp_supercell}) split between $\tilde{k}=\pi$ and $\tilde{k}=0$. Excluding the dispersion origin point leaves seven solutions. (b) Tracking the seven solutions for a range of tuning frequencies of the first resonator $\omega_b$. The tuning frequency of the second resonator is fixed at $\omega_c=1e5$. Solid lines represent $\tilde{k}=\pi$ solutions, dashed-dotted lines represent $\tilde{k}=0$ solutions, and the shaded regions indicate emergent band gaps which are color-coded to indicate one of four scenarios: A - D.} 
	\label{fig:Supercell_omega_c_100000_part1}
\end{figure}

The super cell indicated in Fig.~\ref{fig:Supercell_Sketch_Lumped_masses} consists of two identical outer masses $m_a$ connected in series via a spring of stiffness $k_a$, and two resonators $m_b$ and $m_c$, which are connected to their respective outer masses via the stiffnesses $k_b$ and $k_c$. Following an analysis similar to the one described in Section~\ref{subsection:Dispersion_Analysis_Parallel}, the following dispersion relation can be derived:
\begin{equation}
\begin{bmatrix}
-\omega^2 m_a  + 2 k_a + k_b & -k_a-k_a e^{i\tilde{k}} & -k_b & 0 \\
-k_a - k_a e^{-i\tilde{k}} & - \omega^2 m_a + 2k_a + k_c & 0 & -k_c \\
-k_b & 0 &  - \omega^2 m_b + k_b & 0\\
0 & -k_c & 0 & - \omega^2 m_c + k_c
\end{bmatrix}
\begin{bmatrix}
\tilde{u}_b\\
\tilde{u}_c\\
\tilde{v}_b\\
\tilde{v}_c
\end{bmatrix}
=
\mathbf{0}
\label{eq:disp_supercell}
\end{equation}
where $\tilde{u}_b$, $\tilde{u}_c$, $\tilde{v}_b$, and $\tilde{v}_c$ denote the periodic displacement fields of the first outer mass, second outer mass, first resonator, and second resonator, respectively. Setting $\tilde{k} = 0$ in Eq.~(\ref{eq:disp_supercell}) yields an eighth degree polynomial which has eight roots, four of which are positive, real frequencies. Setting $\tilde{k} = \pi$ in the same equation gives four more real roots for a total of eight solutions. Of these eight solutions, one is guaranteed to be the origin point of the dispersion diagram $[\tilde{k}, \omega] = [0,0]$, as shown in \cite{AlBabaa.2017} (the same point represents a rigid-body mode of a finite, unconstrained metamaterial made up of the same cell). The remaining seven frequency solutions typically represent starting and ending points of dispersion branches (pass bands) with different types of bounded band gaps sandwiched in between. Additionally, the largest of these seven roots is always a part of the $\tilde{k} = 0$ solutions and marks the end of the last dispersion branch and the beginning of the metamaterial's \textit{stop band}, an unbounded band gap which is a hallmark feature of any lumped parameter system \cite{hussein2014dynamics, NouhSinghJASA}. For illustration, an example dispersion diagram showing the aforementioned eight solutions is provided in Fig.~\ref{fig:Supercell_omega_c_100000_part1}a. However, the next subsection will demonstrate that for certain resonator parameters, one or more of these solutions intersect resulting in repeated frequencies, which instigates a number of special scenarios that will be analyzed comprehensively and detailed over the next few figures.

\subsubsection{Shifting, merging, and vanishing of Bragg and local resonance band gaps}
\label{subsection_mergin_supercell}

\begin{figure}[h!]
\centering
	\includegraphics[width=0.98\textwidth]{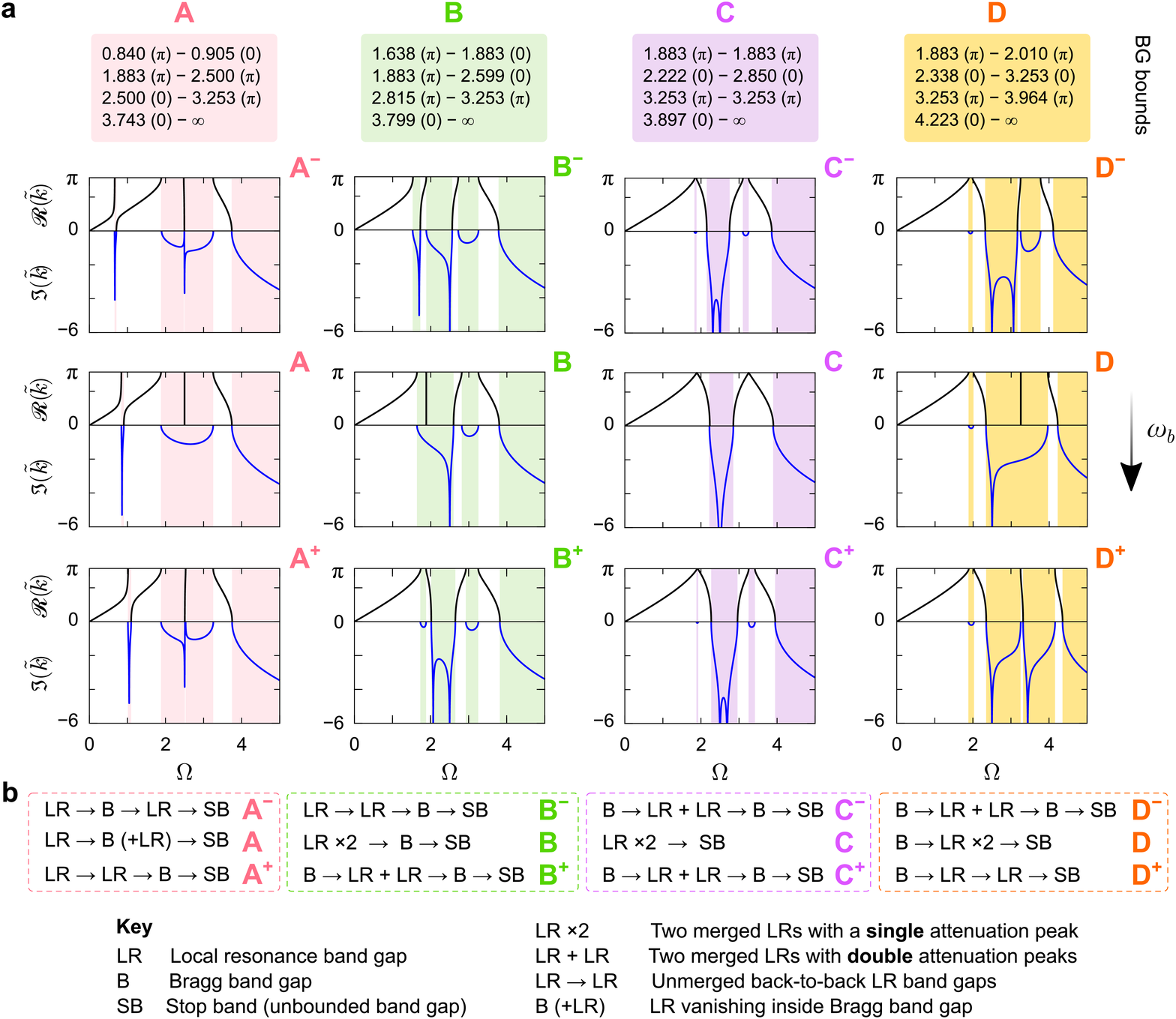}
	\caption{Dispersion scenarios associated with the dual periodic super cell configuration shown in Fig.~\ref{fig:Supercell_Sketch_Lumped_masses}. (a) The top panel lists the seven solutions with the origin of each solution ($\pi$ or $0$) placed between brackets. This list also marks the bounds of the band gap associated with each scenario. The rest of the plot shows three dispersion diagrams corresponding to each of the four scenarios, which show the scenario itself as well as its development using small perturbations of $\omega_b = \pm 0.75e4$. The tuning frequency of the second resonator is fixed at $\omega_c=1e5$. (b) A coded sequence which describes and explains the band gaps taking place in each of the twelve dispersion diagrams. Each band gap type is designated a unique identifier given in the key at the bottom of the figure.} 
	\label{fig:Supercell_omega_c_100000_part2}
\end{figure}

To examine the wide portfolio of dispersion scenarios that can be exhibited by the super cell AM, we start by solving Eq.~(\ref{eq:disp_supercell}) for $m_a=1$ and $k_a=4.8e9$. The tuning frequencies (i.e., stand-alone natural frequencies) of the two resonators are defined as $\omega_b = \sqrt{k_b/m_b}$ and $\omega_c = \sqrt{k_c/m_c}$, which are henceforth used as the main design parameters from which the respective masses and stiffnesses can be computed. We start by fixing $\omega_c$ at a value of $1e5$ and varying $\omega_b$ from $0$ to $2e5$. Figure~\ref{fig:Supercell_omega_c_100000_part1}b illustrates the loci of the seven frequency solutions described earlier as a function of $\omega_b$, for the aforementioned fixed value of $\omega_c$. In the figure, four solid lines indicate the roots corresponding to $\tilde{k}=\pi$ and three dashed-dotted lines indicate the roots corresponding to $\tilde{k}=0$. Given the presence of two different resonators, both of which will be varied in this section, the dimensionless frequency on the vertical axis is henceforth defined as $\Omega = \frac{\omega}{\omega_{b,\text{parallel}}}$, where $\omega_{b,\text{parallel}} = \sqrt{4.8e8/0.3}$ is a constant number that will be fixed throughout. The analysis reveals four $\omega_b$ values of interest which correspond to intersecting solutions, namely: $\omega_{b,\text{A}} \approx 0.34e5$ (Scenario A), $\omega_{b,\text{B}} \approx 0.75e5$ (Scenario B),  $\omega_{b,\text{C}} = 1e5$ (Scenario C), and $\omega_{b,\text{D}} \approx 1.3e5$ (Scenario D). The shaded regions in Fig.~\ref{fig:Supercell_omega_c_100000_part1}b highlight all the band gaps emerging in the four scenarios. While band gap type can usually be inferred from the shape of the dispersion branches (e.g., local resonance gaps are generally characterized by a $\tilde{k}$ phase shift of $\pi$), the same approach cannot be applied here due to the complicated nature of the dispersion diagram and the presence of multiple resonances. Instead, these scenarios can be best understood by examining the full band structures of the super cell obtained using the $\tilde{k}(\Omega)$-solution of the dispersion relation which outputs both the real and imaginary components of the wavenumber, $\Re({\tilde{k}})$ and $\Im({\tilde{k}})$, defining the propagation bands and the attenuation factor, respectively. Figure~\ref{fig:Supercell_omega_c_100000_part2}a shows the complete band structures corresponding to scenarios A through D. The top panel lists the starting and ending frequency of each band gap with the number in brackets indicating whether this frequency represents a $0$ or a $\pi$ solution of $\tilde{k}$. Since these scenarios depict unique special cases, we also provide two additional band structures before and after each scenario which are computed at $\omega_{b,\text{A-D}} \pm 0.75e4$. These explain the development of each scenario and are used to reveal some characteristics of the band gaps which can become hidden when $\omega_b$ exactly matches the value of $\omega_{b,\text{A}}$, $\omega_{b,\text{B}}$, $\omega_{b,\text{C}}$, or $\omega_{b,\text{D}}$.

Upon inspection, the dispersion diagram corresponding to A$^-$ shows a distinct local resonance band gap (LR) at a low frequency (associated with the first resonance $\omega_b$), a stop band (SB) at higher frequencies, and in the middle, a Bragg band gap (B) which is immediately followed by a local resonance band gap (associated with second resonance $\omega_c$). The presence of two local resonance band gaps is expected given the presence of two resonators per super cell. As $\omega_b$ is increased, two features take place simultaneously: one is rather straightforward, while the second is less intuitive but quite interesting. The first feature is a movement to the right of the low-frequency local resonance band gap which is expected since its tuning frequency is directly being varied. Since $\omega_c$ is fixed, the second local resonance band gap does not move and remains at $\omega_c$ (which corresponds to $\Omega=2.5$). However, the second feature associated with $\omega_b$ being varied is that the third dispersion branch flips from being bent upward as $\Re({\tilde{k}})$ goes from $\pi$ to $0$ (in A$^-$), to being bent the opposite way (in A$^+$). The moment when $\omega_b = \omega_{b,\text{A}}$ represents the transition point of this branch and corresponds to the second local resonance vanishing within the Bragg band gap, resulting in Scenario A. At this instant, the third dispersion branch becomes a perfectly flat one at $\Omega=2.5$. In Fig.~\ref{fig:Supercell_omega_c_100000_part1}b, Scenario A can be understood as a result of the intersection of the solid ($\tilde{k}=\pi$) and dashed-dotted ($\tilde{k}=0$) lines at $\omega_{b,\text{A}}$. By purely looking at $\Im({\tilde{k}})$ in the dispersion diagram of A, the system under consideration can be mistakenly interpreted as a traditional single-resonator acoustic metamaterial. However, the seven solutions listed in the figure show two intersecting roots at $\Omega = 2.5$ corresponding to $\tilde{k} = 0$ and $\pi$, which are indicative of an \textit{abrupt} phase shift at the location where the local resonance band gap would have been. The local resonance band gap emerges again as $\omega_b$ is further increased as can be seen in the A$^+$ case.

The diagrams associated with Scenario B show a transition from two back-to-back local resonance band gaps (B$^-$), to a complete merge of the two local resonance band gaps with a single attenuation peak (B), to a merge of the two local resonance band gaps with double attenuation peaks (B$^+$). All three systems exhibit a high-frequency Bragg band gap but only the last is accompanied with an additional low-frequency Bragg band gap which precedes the local resonance merging. Scenarios C$^-$ and C$^+$ show a similar double-peak merging of the local resonance band gap. The Bragg band gap forming before the merge in these two cases becomes very narrow and can only be seen with a fine frequency sampling resolution, or alternatively from the eighth-degree polynomial roots if obtained using high precision. Scenario C depicts a special case where the two resonator tuning frequencies match (i.e., $\omega_b = \omega_c$) resulting in the merged local resonance band gap being the only bounded band gap in the entire dispersion profile, and corresponds to two different intersections of the solid ($\tilde{k}=\pi$) lines at $\omega_{b,\text{C}}$ in Fig.~\ref{fig:Supercell_omega_c_100000_part1}b. Scenario C also represents the traditional single-resonator metamaterial but with a dispersion diagram that is obtained using two rather than one cell. As such, its $\Re({\tilde{k}})$ plot contains only one acoustic branch and one optical branch which appear folded as a result of not using the minimal self-repeating cell. The widest band gap can be witnessed in Scenario D which results from another local resonance band merge with a single attenuation peak. Finally, the case corresponding to D$^+$ shows a very unique design where the \textit{only} Bragg band gap along the frequency axis precedes both local resonance band gaps in an interesting twist which is contrary to traditional metamaterial designs. 

\begin{figure}[h!]
\centering
	\includegraphics[width=0.98\textwidth]{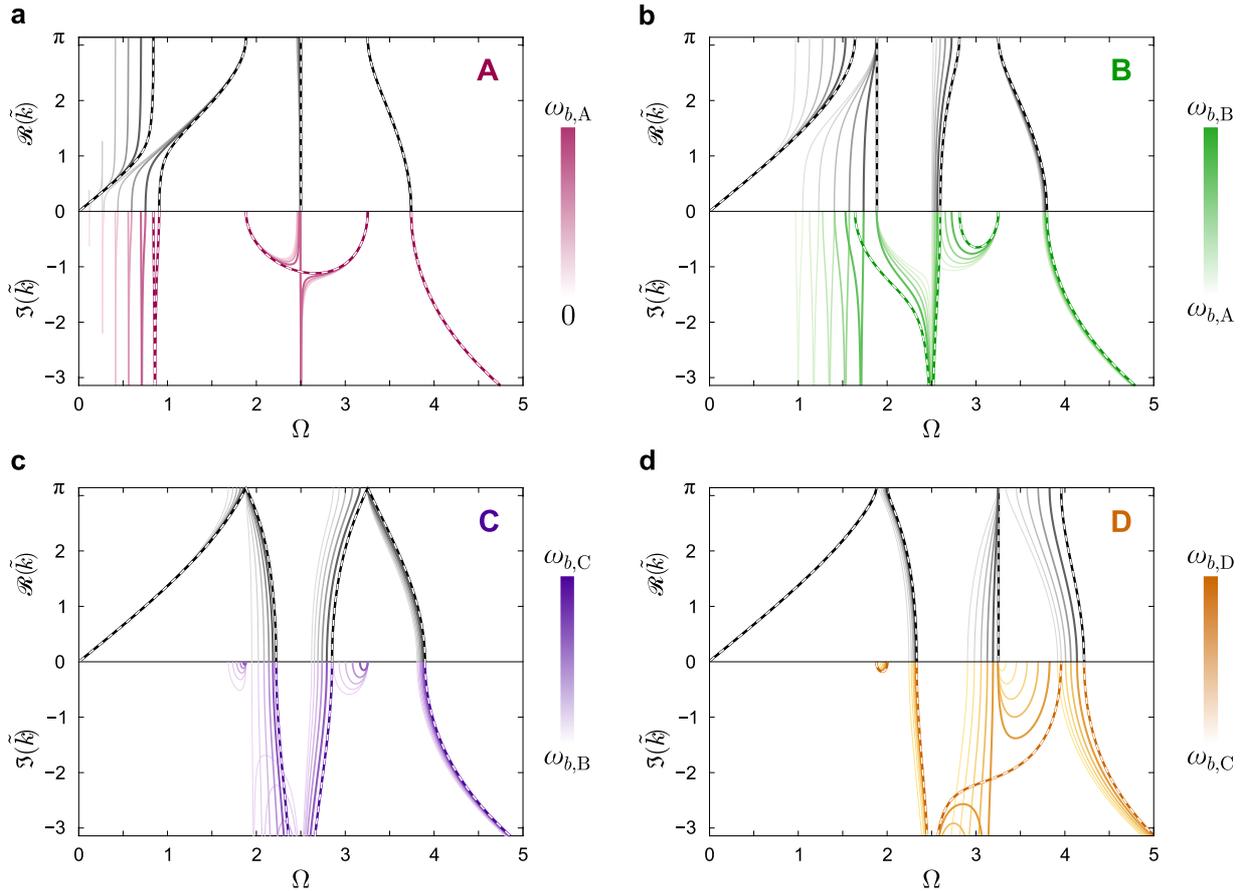}
	\caption{The evolution of the dispersion diagrams associated with the dual periodic super cell configuration shown in Fig.~\ref{fig:Supercell_Sketch_Lumped_masses} for slowly increasing values of $\omega_b$ leading up to (a) Scenario A, (b) Scenario B, (c) Scenario C, and (d) Scenario D. Darker dispersion branches for both $\Re({\tilde{k}})$ and $\Im({\tilde{k}})$ indicate an increasing $\omega_b$.} 
	\label{fig:Supercell_omega_c_100000_gradient}
\end{figure}

The dispersion diagrams presented in Scenarios B$^+$, C$^-$, C$^+$, and D$^-$ are similar to the effect first described in \cite{gao2020ultrawide} (and later exploited in \cite{roca2021multiresonant} and \cite{roca_hussein_dual2021}) which, as can be seen here, represent one specific scenario of many others which can be extracted from the super cell configuration with a smart tuning of the resonant parameters, and that can cater to different applications and frequency regimes. Finally, we also notice that the formation of Bragg band gaps in super cell configurations which exhibit local resonance band gap merging depends on the type of merging, and adheres to the following rule: A Bragg band gap is bound to form both before and after a merged local resonance band gap which exhibits double attenuation peaks (e.g., B$^+$, C$^-$, C$^+$, and D$^-$). However, if the merged local resonance band gap exhibits a single attenuation peak, a Bragg band gap can form before the merge (Case D), after the merge (Case B), or it might not happen at all if $\omega_b = \omega_c$ (Case C). For a complete picture, Fig.~\ref{fig:Supercell_omega_c_100000_gradient} shows the variation of the super cell's dispersion diagram over the entire range of $\omega_b$ frequencies covered by the horizontal axis of Fig.~\ref{fig:Supercell_omega_c_100000_part1}b. An animated illustration of this dispersion diagram evolution can also be seen in the supplementary multimedia file.

\subsubsection{Collapsing solutions and special cases}
\label{subsection_collapsing}

\begin{figure}[h!]
\centering
	\includegraphics[width=0.94\textwidth]{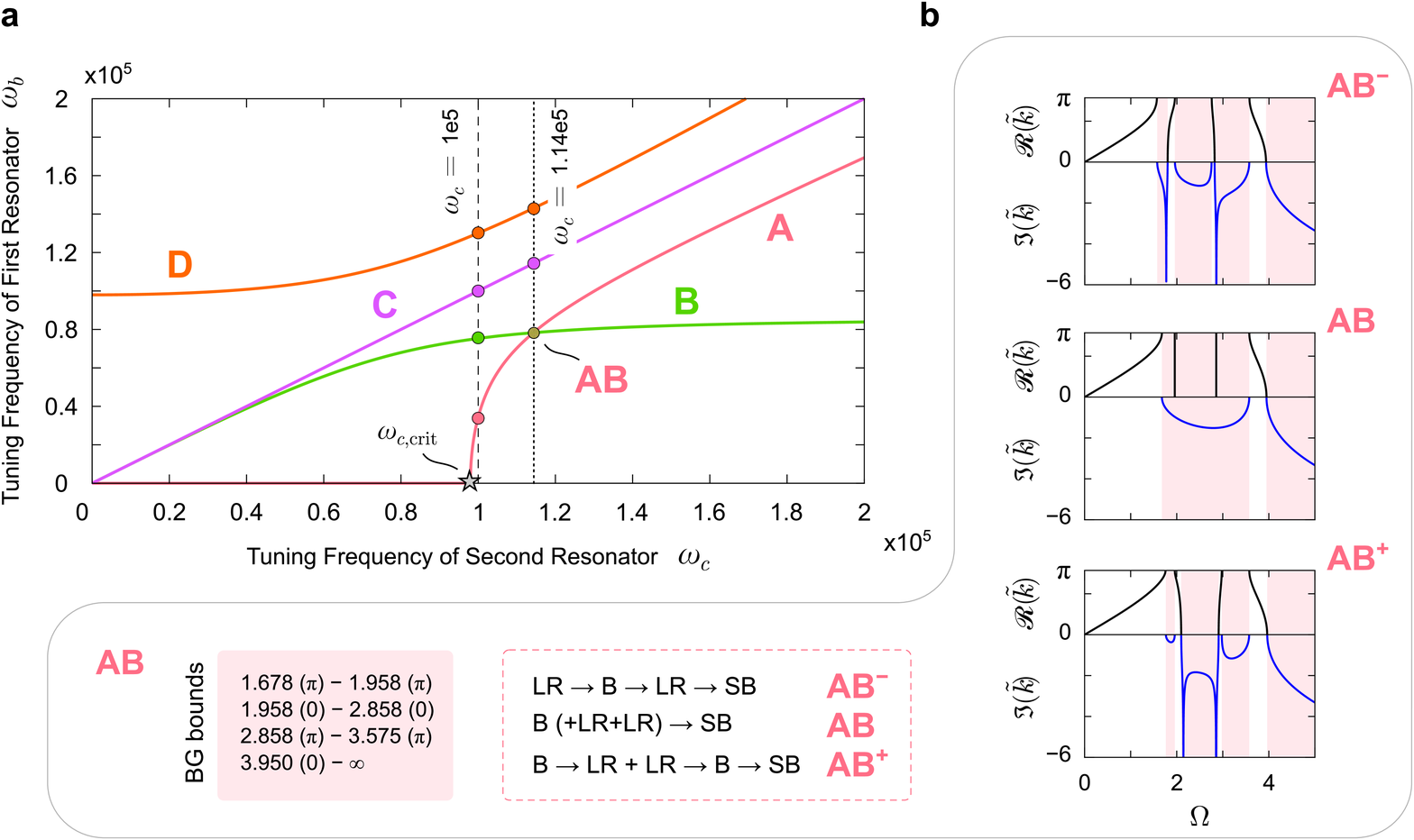}
	\caption{(a) Dispersion scenarios associated with the dual periodic super cell configuration shown in Fig.~\ref{fig:Supercell_Sketch_Lumped_masses} for different combinations of both tuning frequencies $\omega_b$ and $\omega_c$. The configuration labeled with a star marks the first special case $\omega_{c,\text{crit}}$ which corresponds to the vanishing of Scenario A. The second special point is Scenario AB which corresponds to the intersection of the A and B scenarios taking place at $\omega_{b,\text{AB}} = 0.78e4$. (b) Dispersion diagrams and band gap bounds showing the development of Scenario AB.}
	\label{fig:omega_b_over_omega_c}
\end{figure}

If the tuning frequency of the second resonator $\omega_c$ is allowed to change, the wave dispersion scenarios of the super cell illustrated in the previous section can emerge at different combinations of resonator parameters as a result of the larger design space. However, it can also be shown that some of these scenarios disappear, while others converge and eventually coalesce for certain critical combinations. As a case in point, Fig.~\ref{fig:omega_b_over_omega_c}a tracks the four scenarios A through D when any of the $\tilde{k}=\pi$ and the $\tilde{k}=0$ solutions of Eq.~(\ref{eq:disp_supercell}) match (aka collapsing solutions). The vertical dashed line at $\omega_c = 1e5$ shows four distinct collapsing solutions which are portrayed clearly in Fig.~\ref{fig:Supercell_omega_c_100000_part1}. We focus here on two intriguing special cases: The first corresponds to the super cell configuration marked with a ``star" in Fig.~\ref{fig:omega_b_over_omega_c}a, at which point Scenario A ceases to exist. For the values chosen here, this corresponds to $\omega_{c,\text{crit}} = 0.98e4$. For any second resonator tuning frequency below this value (i.e., $\omega_c < \omega_{c,\text{crit}}$), the super cell will continue to show the three scenarios B, C, and D with dispersion behaviors that match those outlined in Section~\ref{subsection_mergin_supercell}. It should also be noted that as for $\omega_c \ll \omega_{c,\text{crit}}$, the B and C scenarios converge and nearly coincide for very low tuning frequencies.  

The second special case pertains to the vertical dotted line at $\omega_c=1.14e5$ which represents a special case where the curves of cases A and B intersect leading to a combined AB scenario. The combined scenario takes place at $\omega_{b,\text{AB}} = 0.78e4$ and culminates in an ultrawide band gap which can be seen in the middle dispersion diagram of Fig.~\ref{fig:omega_b_over_omega_c}b. We refer to this as Scenario AB and for completeness, we provide the dispersion diagrams for the AB$^-$ and AB$^+$ for values of $\omega_b$ slightly before and after $\omega_{b,\text{AB}}$, which show the gradual development of this scenario, in addition to the band gap bounds and the types corresponding to it. Two interesting observations can be made about Scenario AB. The first is that this design exhibits the widest band gap of all the super cell configurations surveyed, extending from $\Omega=1.678$ to $3.575$. The second observation is the lack of a local resonance band gap before the ultrawide band gap. In fact, by interpreting the solutions of the dispersion equation at $\tilde{k}=0$ and $\tilde{k}=\pi$ (listed in Fig.~\ref{fig:omega_b_over_omega_c}b), it can be seen that both local resonances are embedded within the wide band gap resulting in the unique ``B (+LR+LR)" band gap designation indicated in the figure. Once $\omega_b$ is increased past Scenario AB, the super cell exhibits the merged local resonance band gap with double attenuation peaks as confirmed by the AB$^+$ dispersion diagram, with two Bragg gaps before and after and thus adhering to the rule stated earlier in Section~\ref{subsection_mergin_supercell}. Finally, we note that Scenarios C and D  that take place at $\omega_c=1.14e5$ are similar to those outlined in Fig.~\ref{fig:Supercell_omega_c_100000_part2} and show no novel features.

\section{Multi-resonator Acoustic Metamaterials: \textit{Finite Structural Dynamics}}
\label{section:Structural_Dynamics_Analysis}

The features shown thus far have been extracted from the dispersion behavior of the unit cells of the different multi-resonator metamaterials. It is well known that such analysis predicts the wave propagation characteristics of the infinite chain and, while accurate, fails to account for size and boundary effects. The following section of the paper presents a comprehensive mathematical framework which captures the behavior of finite metamaterials made up of both the parallel unit cells and the super cell investigated in Section~\ref{section:Dispersion}.

\subsection{Finite metamaterial with parallel resonators}
\label{subsection:finite_parallel}

Consider a finite acoustic metamaterial with $n$ unit cells and two identical resonators (i.e., $r=2$) per unit cell. The free (unforced) harmonic vibrations of the finite system can be described by:
\begin{equation}
\left(-\Omega^2\omega_b^2
\begin{bmatrix}
\mathbf{M}_a & \mathbf{0} & \mathbf{0}\\
\mathbf{0} & \mathbf{M}_b & \mathbf{0}\\
\mathbf{0} & \mathbf{0} & \mathbf{M}_b
\end{bmatrix}
+ 
\begin{bmatrix}
\mathbf{K}_a + 2\mathbf{K}_b & -\mathbf{K}_b & -\mathbf{K}_b\\
-\mathbf{K}_b & \mathbf{K}_b & \mathbf{0}\\
-\mathbf{K}_b & \mathbf{0} & \mathbf{K}_b
\end{bmatrix}
\right)
\begin{bmatrix}
\mathbf{u}\\
\mathbf{v}_1\\
\mathbf{v}_2
\end{bmatrix}
=
\mathbf{0}
\label{eq:EOMs_r2}
\end{equation}
where $\mathbf{M}_a$, $\mathbf{M}_b$, and $\mathbf{K}_b$ are $n \times n$ diagonal matrices of $m_a$, $m_b$, and $k_b$, respectively, $\mathbf{u} = [u_1\:\: u_2 \:\:... \:\: u_n]^T$, $\mathbf{v}_1 = [v_{1,1}\:\: v_{2,1} \:\:... \:\: v_{n,1}]^T$, and $\mathbf{v}_2 = [v_{1,2}\:\: v_{2,2} \:\:... \:\: v_{n,2}]^T$. Further, $\mathbf{K}_a$ is given by:
\begin{equation}
\underset{n\times n}{\mathbf{K}_{a}} = 
\begin{bmatrix}
k_a & -k_a &  &  & \\
-k_a & 2k_a &  -k_a &  &\\
 & \ddots &  \ddots & \ddots &\\
 &  & -k_a & 2k_a & -k_a\\
 & &  & -k_a & k_a\\
\end{bmatrix}
\end{equation}
Analogous to section~\ref{subsection:Dispersion_Analysis_Parallel}, we generalize Eq.~(\ref{eq:EOMs_r2}) for a finite metamaterial with $n$ unit cells and $r$ resonators per unit cell, which gives:
\begin{equation}
\begin{bmatrix}
-\Omega^2\omega_b^2 \mathbf{M}_a  +\mathbf{K}_a + r\mathbf{K}_b & -\mathbf{K}_b & \cdots & -\mathbf{K}_b\\
-\mathbf{K}_b & -\Omega^2\omega_b^2 \mathbf{M}_b + \mathbf{K}_b & & \\
\vdots & & \ddots & \\
-\mathbf{K}_b & & & -\Omega^2\omega_b^2 \mathbf{M}_b + \mathbf{K}_b
\end{bmatrix}
\begin{bmatrix}
\mathbf{u}\\
\mathbf{v}_1\\
\vdots\\
\mathbf{v}_r
\end{bmatrix}
=
\mathbf{0}
\end{equation}
which can be reduced to:
\begin{equation}
\left(-\Omega^2\omega_b^2 \mathbf{M}_a + \mathbf{K}_a + r\mathbf{K}_b + \left[\Omega^2\omega_b^2 \mathbf{M}_b - \mathbf{K}_b\right]^{-1}r\mathbf{K}_b^2 \right)\mathbf{u} = \mathbf{0} \label{eq:structural_dynamics_compressed_form_generalized}
\end{equation}
For the purposes of this analysis, it is critical that Eq.~(\ref{eq:structural_dynamics_compressed_form_generalized}) be expressed in a perturbed tridiagonal matrix form given by:
\begin{equation}
\begin{bmatrix}
-\eta+b & a & & & \\
a & b &  a & & \\
 & \ddots &  \ddots & \ddots & \\
 & & a & b & a \\
 & & & a & -\epsilon + b
\end{bmatrix}
\mathbf{u}
= 
\mathbf{0} \label{eq:parallel_result_matrix_form_simplified}
\end{equation}
where:
\begin{subequations}
\begin{align}
&a = \Omega^2\omega_b^2k_am_b - k_ak_b \\
&b = \Omega^4\omega_b^4m_am_b - \Omega^2\omega_b^2 m_b(2k_a + rk_b)- \Omega^2\omega_b^2  k_b m_a + 2k_ak_b
\end{align}
\label{eq:structual_dynamics_tridiagonal}
\end{subequations}
and $\eta = \epsilon = -a$ (for the free-free configuration shown in Fig.~\ref{fig:parallel_lumped_mass_system}). The eigenvalues of a matrix of that structure can be analytically obtained for any arbitrary set of parameters using \cite{Yueh.2005}:
\begin{equation}
b+2a\cos\theta_i = 0 \label{eq:lambda_i}
\end{equation}
where $i =1, 2, \dots, n$ and $\theta_i=\frac{i-1}{n}\pi$. In order to extract an estimate for the two frequencies bounding the band gap in the \textit{finite} metamaterial, we  solve for $\Omega$ in Eq.~(\ref{eq:lambda_i}) for specific values of $\theta_i$. Analogous to Eqs.~(\ref{eq:Omega_l_dispersion_analyis_closed_form}) and (\ref{eq:Omega_u_dispersion_analyis_closed_form}) in the dispersion analysis, the lower bound $\Omega_{l,\text{finite}}$ is computed at $\theta_i = \pi$ which can be best approximated by setting $i=n$, which gives:
\begin{align}
\Omega_{l,\text{finite}} =  \frac{1}{\sqrt{2}}\Bigg[2\Gamma &\Big( 1 - \cos \big(\frac{n-1}{n}\pi \big) \Big) + r m_r +  1  \nonumber\\ 
& - \sqrt{\bigg[2\Gamma\Big( 1 - \cos \big(\frac{n-1}{n}\pi \big) \Big) + r m_r +  1\bigg]^2 - 8\Gamma\Big(1- \cos \big(\frac{n-1}{n}\pi \big) \Big)} \hspace{0.1cm} \Bigg]^{1/2}
\label{eq:Omega_l_structural_dynamics_closed_form}
\end{align}
while the upper bound $\Omega_{u,\text{finite}}$ is obtained at $\theta_i = 0$, i.e., $i=1$, giving:
\begin{equation}
\Omega_{u,\text{finite}} = \sqrt{rm_r + 1} 
\label{eq:Omega_u_structural_dynamics_closed_form}
\end{equation}

Eq.~(\ref{eq:Omega_u_structural_dynamics_closed_form}) is identical to dispersion counterpart Eq.~(\ref{eq:Omega_u_dispersion_analyis_closed_form}), and shows that the upper band gap bound will remain at the same frequency irrespective of the number of unit cells $n$ in the finite metamaterial chain. Even at the finite level, $\Omega_{u,\text{finite}}$ remains only dependent on the number of resonators $r$ and on the mass ratio $m_r$ defined earlier. A similar observation was made in \cite{AlBabaa.2017} for a single-resonator metamaterial. Eq.~(\ref{eq:Omega_l_structural_dynamics_closed_form}), on the other hand, shows that the lower band gap is dynamic and changes as the number of unit cells $n$ increases, in addition to being influenced by the number of resonators $r$ and both the stiffness and mass ratios. However, it is important to note that at the infinite limit $n \to \infty$, the cosine term approaches a value of $-1$ rendering $\Omega_{l,\text{finite}} = \Omega_l$ as expected and bridging the two methodologies detailed in sections~\ref{subsection:Dispersion_Analysis_Parallel} and \ref{subsection:finite_parallel}.

\subsubsection{Transfer Function Approach}
\label{subsection:Transfer_Function_Approach}

\begin{figure}[h!]
\centering
	\includegraphics[width=0.85\textwidth]{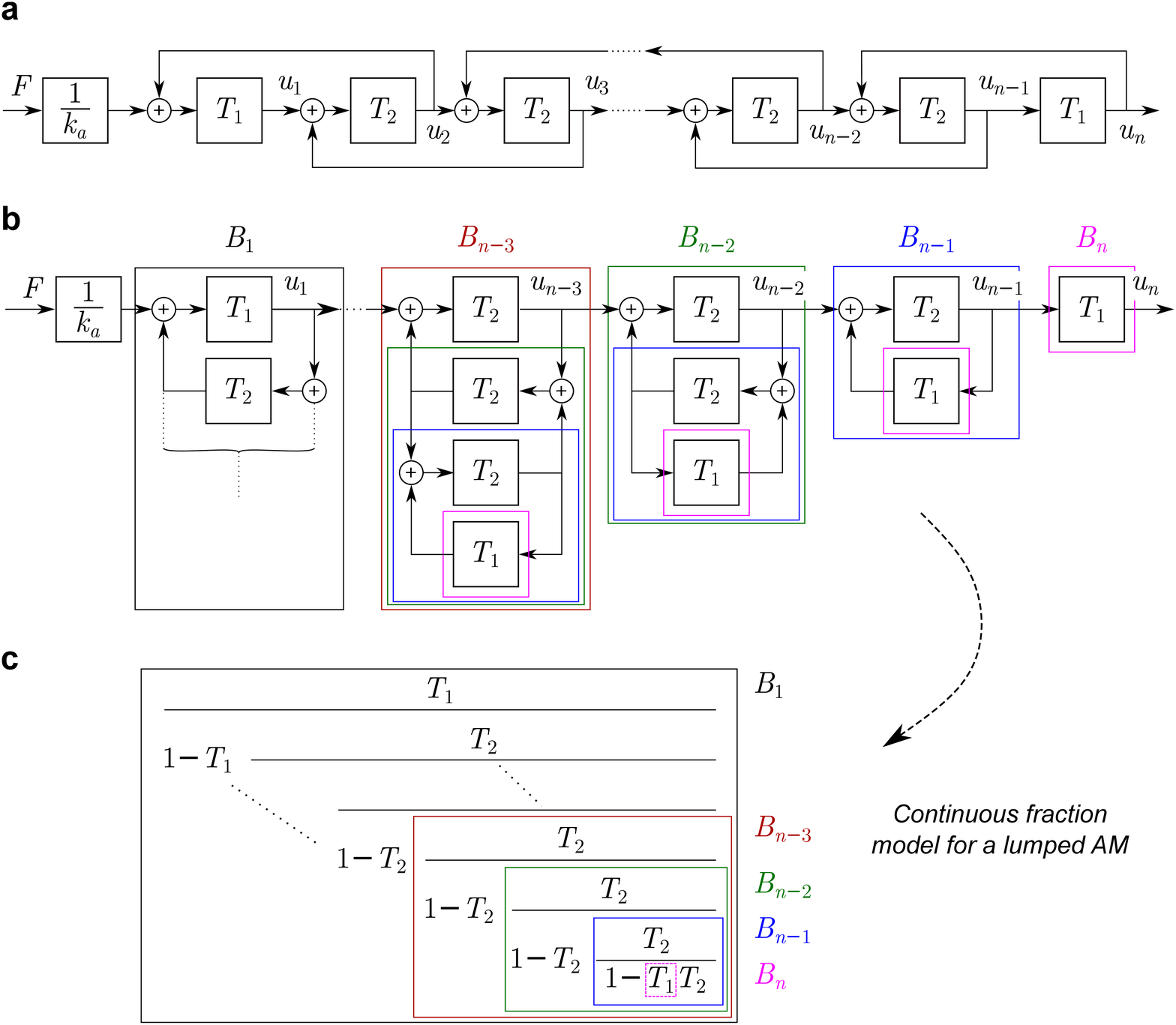}
	\caption{Simplification of the transfer function model of a finite metamaterial via block diagram reduction. \label{fig:TF_approach}}
\end{figure}
\noindent

For a metamaterial with $n$ unit cells and $r$ identical resonators per unit cell, which is subject to an excitation force acting on the outer mass of the first cell (see Fig.~\ref{fig:parallel_lumped_mass_system}), we place a force $F$ in the first element of the column vector on the right side of Eq.~(\ref{eq:EOMs_r2}). The dynamics of the finite chain can be described by the block diagram shown in Fig.~\ref{fig:TF_approach}a, where:
\begin{equation}
T_1(s) = \frac{k_a(m_bs^2 + k_b)}{(m_as^2 + k_a+rk_b)(m_bs^2 + k_b)-rk_b^2} \label{eq:T1}
\end{equation}
and $s$ is a Laplace complex number. The transfer function $T_1(s)$ is unique to the beginning and end of the chain, i.e., the $1^{\text{st}}$ and $n^{\text{th}}$ unit cells, respectively. Input-output relations for the rest of the chain are given by $T_2(s)$:
\begin{equation}
T_2(s) = \frac{k_a(m_bs^2 + k_b)}{(m_as^2 + 2k_a+rk_b)(m_bs^2 + k_b)-rk_b^2} \label{eq:T2}
\end{equation}

A manipulation of this block diagram results in the structure shown in Fig.~\ref{fig:TF_approach}c. As can be seen in the figure, every subsystem can be regarded as a positive feedback loop. Furthermore, the following observations can be made:
\begin{subequations}
\begin{align}
&B_n = T_1 \label{eq.:B_n}\\
&B_m = \frac{T_2}{1-T_2B_{m+1}}\label{eq.:B_m}\\
&B_1 = \frac{T_1}{1-T_1B_2}\label{eq.:B_1}
\end{align}
\end{subequations}
where $2 \leq m \leq n-1$ and Eq.~(\ref{eq.:B_m}) is only valid for $n \geq 3$. The transfer function relating the displacement of the $i^{\text{th}}$ cell to the incident force is therefore given by:
\begin{equation}
\frac{u_i}{F} = \frac{1}{k_a}\prod_{j=1}^{i}B_j
\end{equation}
where $i > j$. The $T_1$ term in Eq.~(\ref{eq.:B_1}) can be thought of as the numerator of the continued fraction illustrated in Fig.~\ref{fig:TF_approach}c. Consider, for example, a metamaterial with $n=2$ cells and $r$ identical resonators. In such a case, $B_n = B_2 = T_1$ and $B_1 = \frac{T_1}{1-T_1^2}$, and the system then has the end-to-end transfer function:
\begin{equation}
\frac{u_2}{F} = \frac{1}{k_a}B_1B_2 = \frac{1}{k_a} \frac{T_1^2}{1-T_1^2}\label{eq:u2_F}
\end{equation}
The poles of this transfer function are the roots of:
\begin{equation}
\big[(m_as^2 + k_a+rk_b)(m_bs^2 + k_b)-rk_b^2\big]^2 -\big[k_a(m_bs^2 + k_b)\big]^2 = 0
\end{equation}
from which, upon simplification, the following can be factored out:
\begin{equation}
m_am_bs^2+(m_a+rm_b)k_b=0
\label{eq:factored}
\end{equation}

Making use of $s=j\omega=j\Omega\sqrt{k_b/m_b}$, where $j = \sqrt{-1}$, Eq.~(\ref{eq:factored}) boils down to $\Omega = \sqrt{rm_r+1}$ representing the system pole flanking the band gap at the high frequency side, i.e., the upper bound $\Omega_u$, and thus validating this theoretical framework.

\subsubsection{Band gap widening with overall size/mass reduction}
\label{subsection:BG_widening_finite}
\begin{figure}[h!]
\centering
	\includegraphics[width=\textwidth]{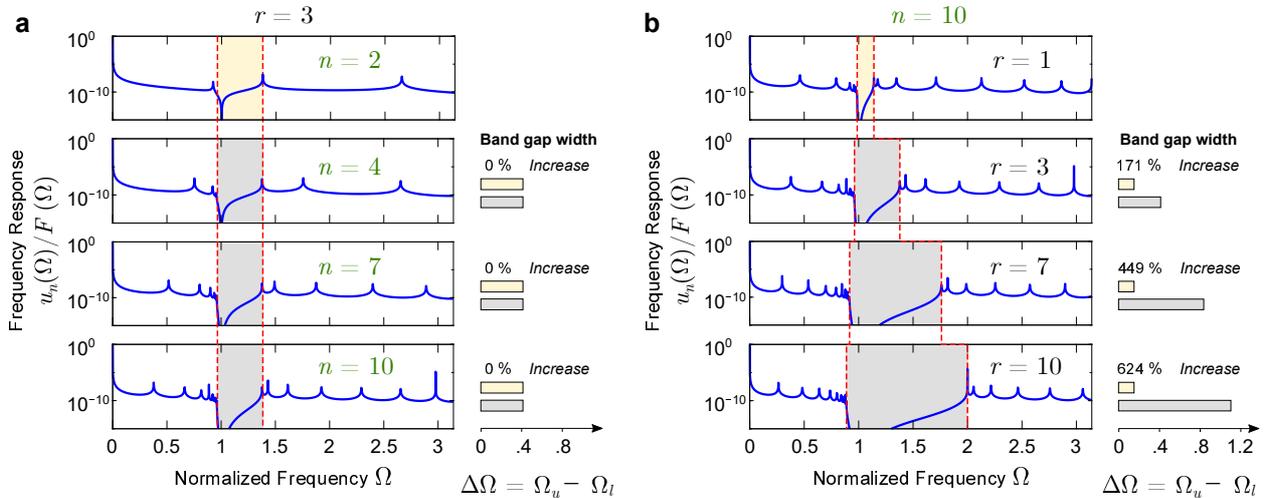}
	\caption{Frequency response of the $n^{\text{th}}$ outer mass displacement of a finite acoustic metamaterial with (a) $r=3$ parallel resonators and $n=2, 4, 7$, and $10$ cells, and (b) $n=10$ cells and $r=1, 3, 7$, and $10$ parallel resonators per cell as a ratio of an excitation applied at the opposite end. Shaded regions indicate band gap obtained via dispersion analysis. Adjacent plots show increase in band gap width $\Delta \Omega$ for the various cases.}
	\label{fig:n_versus_r_effect}
\end{figure}

In this section, the performance of a finite lumped acoustic metamaterial with parallel identical resonators is evaluated. The AM parameters are kept as follows: $m_a=1$, $m_b=0.3$, $k_a=4.8e9$, and $k_b=0.1k_a$ to ensure a fair comparison with the results obtained earlier from the unit cell analysis. We begin by confirming that for the same number of resonators per unit cell, increasing the number of cells of the finite AM does not fundamentally change the size of the theoretically predicted band gap. Instead, the band gap gradually takes shape and eventually converges to the bounds predicted by the dispersion analysis. Fig.~\ref{fig:n_versus_r_effect}a shows the frequency response function $u_n/F$, which depicts the end displacement of the AM $u_n$ as a ratio of an incident force $F$ at the opposite end for an AM with $r=3$ parallel resonators per unit cell, for a finite chain comprising $n=2$, $4$, $7$, and $10$ cells, respectively. As the number of cells increases, the lower band gap bound of the finite metamaterial, i.e., $\Omega_{l,\text{finite}}$ approaches the dispersion lower bound $\Omega_l$, while the upper bound remains unchanged ($\Omega_{u,\text{finite}} = \Omega_u$). For this example, the lower bounds of the finite metamaterial are found to be $\Omega_{l,\text{finite}}=0.9226$,
$0.9550$, $0.9597$, and $0.9607$ for $n=2$, $4$, $7$, and $10$, respectively, which correspond to an $\Omega_l$ value of $0.9617$ for the infinite system. These values can be confirmed using Eqs.~(\ref{eq:Omega_l_dispersion_analyis_closed_form}) and (\ref{eq:Omega_l_structural_dynamics_closed_form}). The shaded regions in the shown plots represent the band gap bounds as predicted by the dispersion analysis for an infinite chain, which serves as a reference for the finite analysis.

Unlike Fig.~\ref{fig:n_versus_r_effect}a where the number of resonators was kept constant, Fig.~\ref{fig:n_versus_r_effect}b shows the effect of increasing the number of ``parallel" resonators per each cell from $1$ to $10$ on the same frequency response function. In this scenario, all the finite metamaterials consist of $n=10$ cells to ensure a fair comparison. Here, the potential of the parallel multi-resonator design becomes clearly evident, as the band gap width significantly widens with an increase in $r$. Similar to Fig.~\ref{fig:n_versus_r_effect}a, the blue curves are a result of the transfer function approach for the finite structure (section~\ref{subsection:Transfer_Function_Approach}), while the shaded regions represent the band gap bounds obtained from the dispersion analysis (section~\ref{subsection:Dispersion_Analysis_Parallel}). Since the number of cells $n$ is sufficiently large, the shaded regions align well with the frequency response. More importantly, the dotted red lines, which trace the band gap bounds, show that the increase in $r$ widens the band gap from both ends, albeit more significantly at the upper bound frequency. For this example, the lower band gap bounds of the finite AM are found to be $\Omega_{l,\text{finite}}=0.9863$, $0.9607$, $0.9155$, and $0.8857$ for $r=1$, $3$, $7$, and $10$, respectively, which correspond to the following dispersion values: $\Omega_{l,r=1}=0.9867$, $\Omega_{l,r=3}=0.9617
$, $\Omega_{l,r=7}=0.9174$, and $\Omega_{l,r=10}=0.8882$. While the lower bound decreases with an increase in $r$, it is worth noting that such lower bound is always smaller for a finite AM than for an infinite one, i.e., $\Omega_{l,\text{finite}} < \Omega_l \hspace{0.2cm} \forall \hspace{0.2cm} r$. The upper band gap bounds are $\Omega_{u,r=1}=1.1402$, $\Omega_{u,r=3}=1.3784
$, $\Omega_{u,r=7}=1.7607$, and $\Omega_{u,r=10}=2$, which can also be confirmed using either Eq.~(\ref{eq:Omega_u_dispersion_analyis_closed_form}) or (\ref{eq:Omega_u_structural_dynamics_closed_form}).

To better quantify the improvement in band gap size made possible by the parallel assembly of identical resonators, let's define the band gap width as $\Delta\Omega=\Omega_u-\Omega_l$, which can be used as a comparative metric for different AMs. Using the single-resonator AM (i.e., $r=1$) as a benchmark, a multi-resonator AM exhibits an increase of $171\%$, $449 \%$, and $624 \%$ in $\Delta \Omega$ using $r=3$, $7$, and $10$, respectively, as shown in the rightmost panel of Fig.~\ref{fig:n_versus_r_effect}b. Perhaps even more importantly, as seen in Fig.~\ref{fig:n_versus_r_effect}b, such increase in band gap width in finite AMs of the same number of cells takes place without adding resonant peaks to the frequency response.

\begin{figure}[h!]
\centering
	\includegraphics[width=\textwidth]{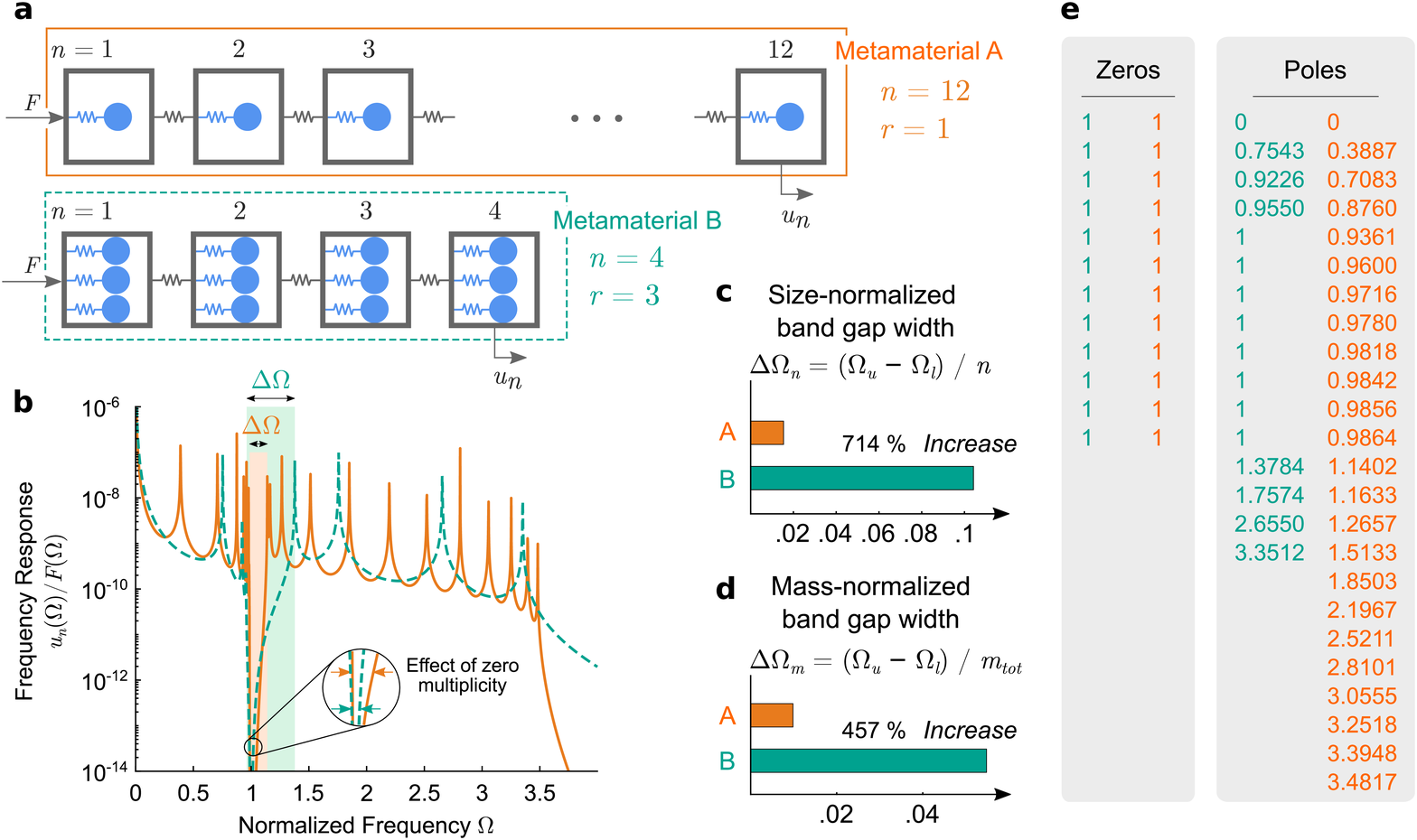}
	\caption{Comparison between ``Metamaterial A" with $n=12$ single-resonator cells and ``Metamaterial B" with $n=4$ cells and $3$ resonators in each: (a-b) Schematic diagram and frequency response functions of both metamaterials. (c-d) Comparison of the size-normalized and mass-normalized band gap widths, $\Delta \Omega_n$ and $\Delta \Omega_m$, respectively. (e) A list of the transfer function poles and zeros of both metamaterials showing significantly fewer resonant peaks in the response of B than A over the entire frequency spectrum covered by both systems.}
\label{fig:BG_comparision_n12_r1_n4_r3}
\end{figure}

As noted in Section~\ref{subsection:Dispersion_Analysis_Parallel}, the band gap widening effect associated with parallel resonators requires an increase in unit cell mass. However, this does not necessarily translate to an overall mass increase of the finite metamaterial. This stems from the fact that the overall number of resonators in a finite metamaterial plays the most significant role in the resultant band gap size. As such, identical resonators which are stacked in parallel and housed in a fewer number of cells can result in a compact metamaterial which exhibits a stronger band gap with a smaller footprint. As a case in point, Fig.~\ref{fig:BG_comparision_n12_r1_n4_r3} pits two finite AMs against each other. In this example, Metamaterial A is a traditional single-resonator system comprised of 12 cells, while Metamaterial B is made out of 4 cells only with 3 parallel resonators per cell. All the parameters of the outer as well as resonant spring-mass systems are kept the same as those used in Fig.~\ref{fig:n_versus_r_effect}. As a result, the two systems include the same total number of resonators ($r_{tot}=nr=12$). However, Metamaterial B is overall lighter than A since it comprises fewer outer masses, i.e., $m_{tot,\text{B}} < m_{tot,\text{A}}$. We define a size-normalized band gap width metric $\Delta \Omega_n = (\Omega_u - \Omega_l)/n$, which weighs in the length of the lumped chain in comparing the width of band gaps in the two finite AMs, as well as a mass-normalized band gap width metric $\Delta \Omega_m = (\Omega_u - \Omega_l)/m_{tot}$, where $m_{tot} = n (m_a + r m_b)$, which weighs in the total mass of the lumped chain in comparing both systems. As seen in the figure, the multi-resonator AM response is superior on three fronts:

\begin{enumerate}
    \item It exhibits a $714 \%$ increase in $\Delta \Omega_n$ over a traditional single-resonator metamaterial, thus producing a wider band gap with less than half the number of cells.
    \item It exhibits a $457 \%$ increase in $\Delta \Omega_m$ over a traditional single-resonator metamaterial, thus producing a wider band gap with a reduction in overall mass.
    \item It exhibits less than half the resonant frequencies of the traditional single-resonator metamaterial (specifically 8 versus 24 peaks), since the additional outer masses in the $n=12$ case add new (non-repeated) poles to the frequency spectrum.
\end{enumerate}

The third point is of further interest since it explains the band gap widening phenomenon associated with parallel identical resonators, but rather from the perspective of the finite analysis. As outlined in Ref.~\cite{AlBabaa.2017}, the formation mechanism of a local resonance band gap in a finite metamaterial is a byproduct of two features: The first is a multiplicity of repeated transfer function \textit{zeros} at the tuning resonator frequency (i.e., $\Omega =1$). This multiplicity tends to flatten out the frequency response and convert it from a single notch to a flatter bowl-shaped curve at low vibration amplitudes. The second is the presence of two transfer function \textit{poles} which culminate in two resonant peaks that represent the beginning and end of the band gap. Figure~\ref{fig:BG_comparision_n12_r1_n4_r3}e lists the poles and zeros of the response plotted in Fig.~\ref{fig:BG_comparision_n12_r1_n4_r3}b for both metamaterials, A and B. The results show that the band gap associated with Metamaterial B spans a wider frequency range because it exhibits a lower band gap starting pole (0.955 vs. 0.986) and a higher band gap ending pole (1.378 vs. 1.140) which is consistent with the calculations of $\Omega_{l,\text{finite}}$ and $\Omega_{u,\text{finite}}$, respectively. However, they also show that although narrower towards higher amplitudes, the band gap exhibited by Metamaterial A remains wider at very low amplitudes, as evident in the close-up inset of Fig.~\ref{fig:BG_comparision_n12_r1_n4_r3}b. This is a direct consequence of the zero multiplicity effect stated earlier. More specifically, as listed in the figure, although both systems exhibit 12 repeated zeros at $\Omega =1$ caused by the 12 individual resonators, the transfer function of Metamaterial B generates 8 repeated poles at $\Omega=1$ which cancel out and mildly weaken the effect of the repeated zeros in the parallel design. This tends to happen at extremely low amplitudes which are practically immeasurable, rendering the parallel band gap significantly advantageous for the criteria mentioned above. 

\begin{figure}[h!]
\centering
	\includegraphics[width=\textwidth]{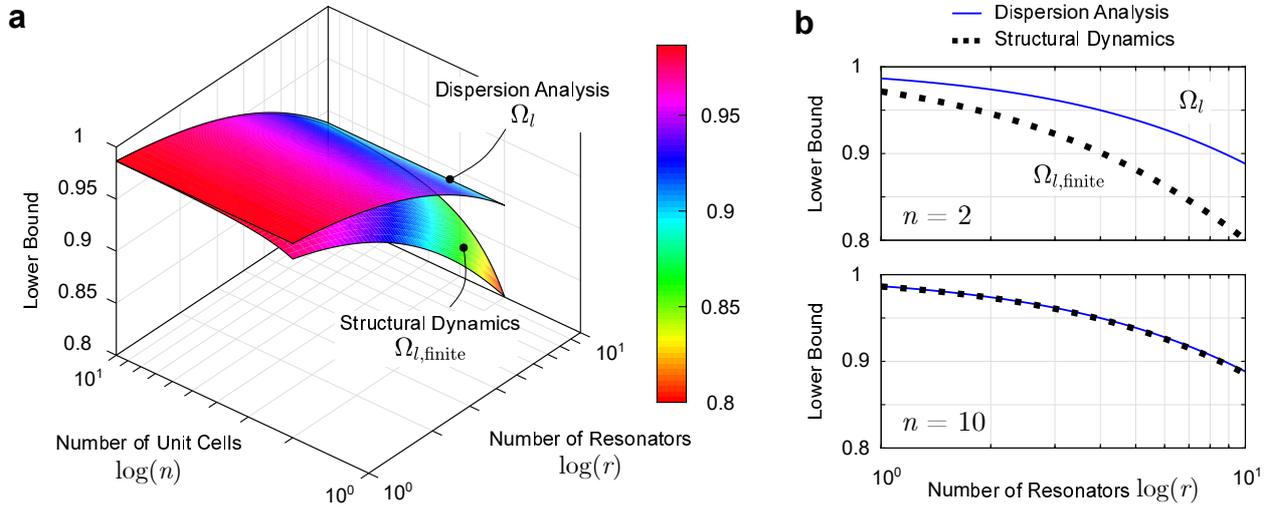}
	\caption{(a) Surface plot showing the lower band gap bound dependence on the number of resonators $r$ and the number of cells $n$ using the unit cell dispersion analysis (i.e., $\Omega_l$) and the finite structural dynamics approach (i.e., $\Omega_{l,\text{finite}}$). (b) Lower band gap bound versus $r$ for $n=2$ and $n=10$, showing convergence between the two approaches at sufficiently large values of $n$.}
	\label{fig:plot_surface_lower_bound_final}
\end{figure}

While the ability to house several resonators within the same host cell depends on metamaterial type and space constraints, it is easy to conceive of a metamaterial design with three identical resonators, especially given the advent of additive and high-precision manufacturing which can cater to applications with both size and weight constraints. Finally, Fig.~\ref{fig:plot_surface_lower_bound_final}a provides an illustrative design chart which summarizes the variation of the lower band gap bound with a varying number of unit cells $n$ and resonators $r$. The surface plot in Fig.~\ref{fig:plot_surface_lower_bound_final}a depicts the difference between $\Omega_l$ and $\Omega_{l,\text{finite}}$ explained earlier and shows that for high values of $r$ and small values of $n$, the discrepancy between these two values is maximized. For AMs with a sufficiently large number of cells $n$, the two approaches converge to the same value regardless of the number of resonators; an expected feature as the finite systems approaches the infinite limit. The latter can be clearly observed in Fig.~\ref{fig:plot_surface_lower_bound_final}b.

\subsubsection{Hybrid parallel-series configuration}
\label{subsection:hybrid_parallel_series}

 \begin{figure}[h!]
 \centering
 	\includegraphics[width=0.97\textwidth]{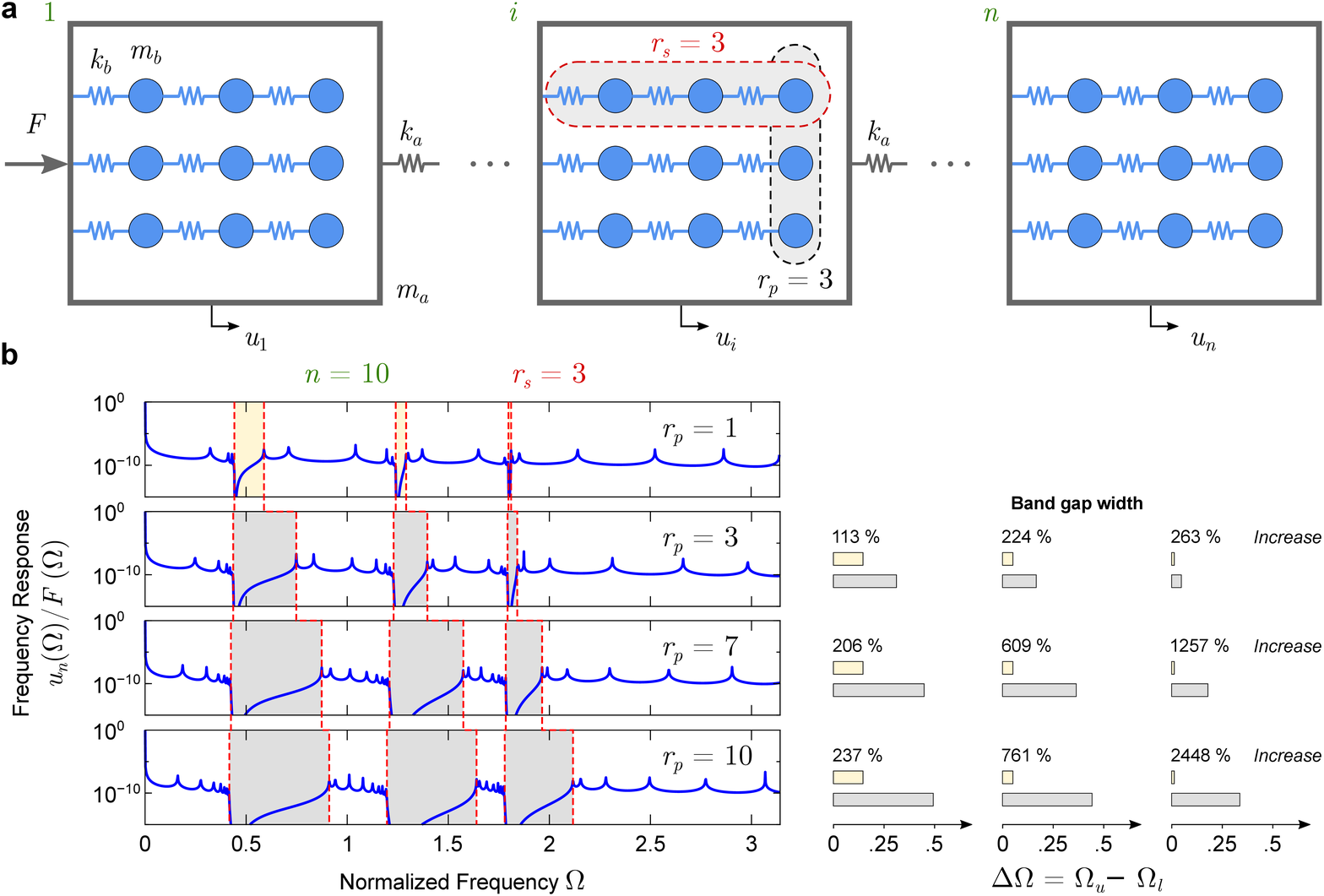}
 	\caption{(a) Lumped parameter model of a multi-resonator acoustic metamaterial with hybrid resonator arrangement. (b) Frequency response of the $n^{\text{th}}$ outer mass displacement of a finite acoustic metamaterial with $n=10$, $r_s=3$ resonators in series, and $r_p=1$, $3$, $7$, and $10$ rows in parallel. Adjacent plots show increase in band gap width $\Delta \Omega$ over a single-resonator design for the various cases shown.}
  \label{fig:n_versus_r_hybrid}
 \end{figure}

While parallel resonators can be used to widen local resonance band gaps, a hybrid parallel-series unit cell configuration can be used to selectively place band gaps at different frequencies, while being able to widen/enhance these multiple gaps at the same time. In such a design, the number of distinct band gaps is equal to the number of in-series resonators, referred to henceforth as $r_s$ and the widening effect remains a function of the number of in-parallel resonators, referred to henceforth as $r_p$. As a result, the total number of resonators $r$ housed in each unit cell of the hybrid design can be determined from $r = r_s r_p$, and the overall number of resonators placed overall in a finite metamaterial becomes equal to $n r$. A schematic diagram of the hybrid design is shown in Fig.~\ref{fig:n_versus_r_hybrid}a. A finite hybrid multi-resonator metamaterial is constructed using the same values of $m_a$, $m_b$, $k_a$, and $k_b$ used throughout this paper. In other words, all the resonators in the hybrid cell are kept identical, with the understanding that an optimization of such values can be later used to change the location of the band gaps. In this example, we choose a metamaterial which consists $r_s=3$ in-series resonators, and four hybrid designs which correspond to $r_p=1$, $3$, $7$, and $10$ parallel repetitions of this set of in-series resonators. Stated differently, the four unit cell designs are comprised of $r=3$, $9$, $21$, and $30$ resonators, respectively. All the finite chains consist of $n=10$ cells. Fig.~\ref{fig:n_versus_r_hybrid}b shows the same frequency response function $u_n/F$ for the four aforementioned designs. The top plot corresponds to $r_s=3$ and $r_p=1$, which is simply three resonators in series. The presence of three resonators in series results in the emergence of the three band gaps that are shown in the figure. The second, third, and fourth plots of Fig.~\ref{fig:n_versus_r_hybrid}b show how each and every one of these band gaps can be widened by repeating the in-series resonators several times in parallel. Table~\ref{table:hybrid} lists the lower and upper band gap bounds, $\Omega_l$ and $\Omega_u$, for all four cases as obtained from a unit cell dispersion analysis. Consistent with our previous analysis, a gradual (small) decrease in $\Omega_l$ and a gradual (large) increase in $\Omega_u$ take place as $r$ increases. Interestingly, Fig.~\ref{fig:n_versus_r_hybrid}b also reveals that while the increase in the first band gap width seems to slow down or saturate with the growing number of parallel arrangements in the hybrid design, the third band gap continues to significantly expand. This could be attributed to the presence of fewer constraints (e.g., additional band gaps) on the high frequency side of the third and last band gap in this particular design. Nonetheless, the bottom plot of Fig.~\ref{fig:n_versus_r_hybrid}b shows prodigious percentage increases of $237 \%$, $761 \%$, and $2,448 \%$ in the band gap width $\Delta \Omega$ over the single-resonator design, which is a testament to the potential of intelligently designed multi-resonator acoustic metamaterials to provide both strong and broadband vibroacoustic suppression. 

\begin{table}[h!]
\caption{Lower and upper band gap bounds, $\Omega_l$ and $\Omega_u$, for a hybrid multi-resonator cell with $r_s=3$.}
\centering 
\small
\begin{tabular}{l c c c c c c} 
\toprule 
 & Band gap & $r_p=1$ & $r_p=3$ & $r_p=7$ & $r_p=10$\\
 \midrule
 \multirow{3}{*}{$\Omega_l$} & 1 & $0.4420$ & $0.4361$ & $0.4248$ & $0.4169$\\
 & 2 & $1.2409$ & $1.2295$ & $1.2095$ & $1.1967$\\
 & 3 & $1.7988$ & $1.7932$ & $1.7844$ & $1.7794$\\
 \midrule
 \multirow{3}{*}{$\Omega_u$} & 1 & $0.5886$ & $0.7481$ & $0.8737$ & $0.9110$\\
 & 2 & $1.2923$ & $1.3964$ & $1.5744$ & $1.6398$\\
 & 3 & $1.8120$ & $1.8413$ & $1.9641$ & $2.1169$\\
 \bottomrule
 \label{table:hybrid}
 \end{tabular}
 \end{table}

\subsection{Finite metamaterial with super cell configuration}
\label{subsection:Transfer_Function_Approach_supercell}

\begin{figure}[h!]
\centering
	\includegraphics[width=0.85\textwidth]{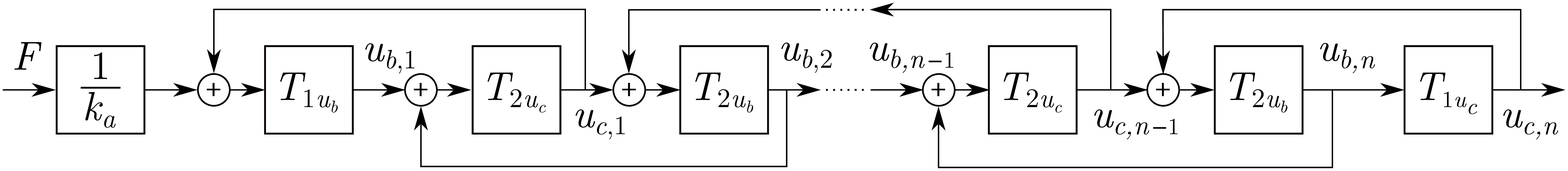}
	\caption{Simplification of the transfer function approach for the super cell \label{fig:TF_final_supercell}}
\end{figure}

A finite metamaterial with $n$ super cells is considered, where each super cell has the parameters defined in Fig.~\ref{fig:Supercell_Sketch_Lumped_masses}. The dynamics of the finite chain can be described by Fig.~\ref{fig:TF_final_supercell} where:
\begin{subequations}
\begin{equation}
T_{1u_{b,c}}(s) = \frac{k_a}{m_as^2 + k_a + k_{b,c} - \frac{k_{b,c}^2}{m_{b,c}s^2+k_{b,c}}}
\end{equation}
\begin{equation}
T_{2u_{b,c}}(s) = \frac{k_a}{m_as^2 + 2k_a + k_{b,c} - \frac{k_{b,c}^2}{m_{b,c}s^2+k_{b,c}}}
\end{equation}
\end{subequations}

\noindent

The transfer functions $T_{1u_b}(s)$ and $T_{1u_c}(s)$ are unique to the first and last super cells of the finite chain, where input-output relations for the rest of the chain are given by $T_{2u_b}(s)$ and $T_{2u_c}(s)$. As shown in Fig.~\ref{fig:TF_final_supercell}, every subsystem can be regarded as a positive feedback loop and, therefore, the following block diagram simplifications can be made:
\begin{subequations}
\begin{align}
&B_{2n} = T_{1u_c}\label{eq.:B_n_supercell}\\
&B_m = \frac{T_{2u_b}}{1-T_{2u_b}B_{m+1}} \hspace{0.8cm} \text{(for odd $m$)}\label{eq.:B_m_supercell}\\
&B_m = \frac{T_{2u_c}}{1-T_{2u_c}B_{m+1}} \hspace{1cm} \text{(for even $m$)}\label{eq.:B_m_supercell}\\
&B_1 = \frac{T_{1u_b}}{1-T_{1u_b}B_2}\label{eq.:B_1_supercell}
\end{align}
\end{subequations}
where $2 \leq m \leq 2n-1$ and Eq.~(\ref{eq.:B_n_supercell})-(\ref{eq.:B_1_supercell}) are only valid for $n \geq 2$. The transfer functions relating the displacement of the different masses of the $i^{\text{th}}$ super cell to the incident force are therefore given by:
\begin{subequations}
\begin{align}
&\frac{u_{b,i}}{F} = \frac{1}{k_a}\prod_{j=1}^{2i-1}B_j  \hspace{1.65cm} (\text{for $i>j$})\\
&\frac{u_{c,i}}{F} = \frac{1}{k_a}\prod_{j=1}^{2i}B_j  \hspace{1.7cm} (\text{for $i>j$})\\
&\frac{v_{b,i}}{F} = \frac{k_b}{s^2m_b+k_b}u_{b,i}
\hspace{1cm} (\text{for $i>j$})\\
&\frac{v_{c,i}}{F} = \frac{k_c}{s^2m_c+k_c}u_{c,i}
\hspace{1cm} (\text{for $i>j$})\label{eq:B_j_supercell}
\end{align}
\end{subequations}

\begin{figure}[h!]
\centering
	\includegraphics[width=\textwidth]{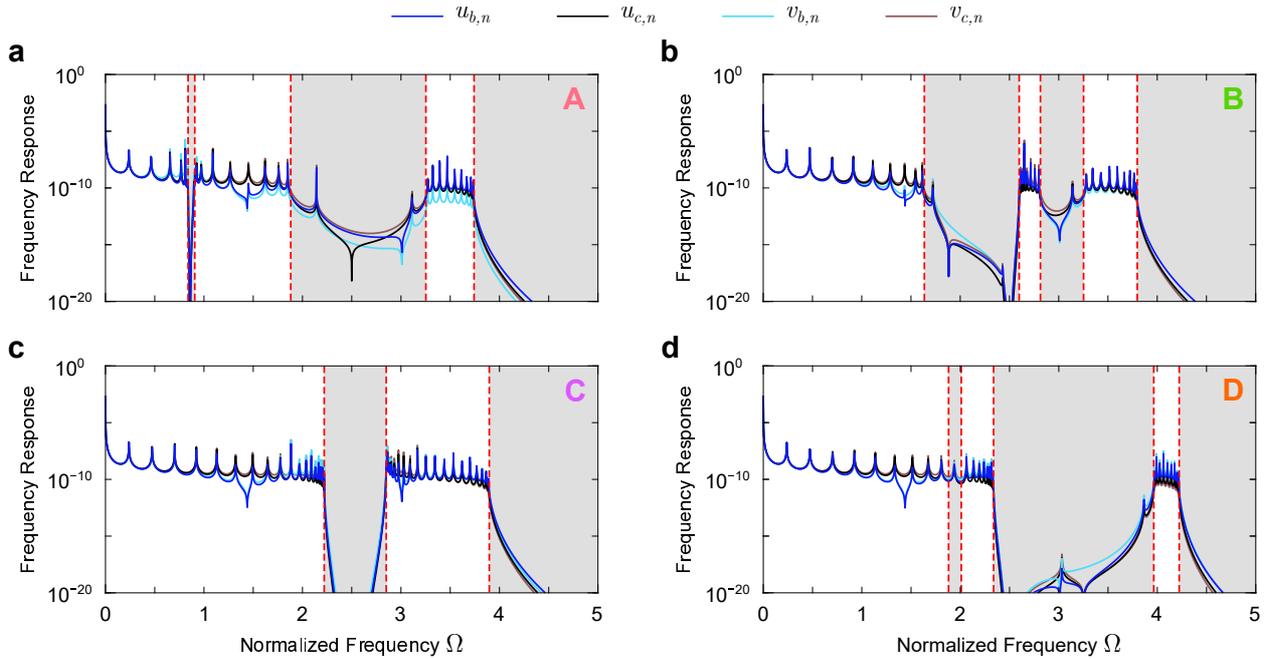}
	\caption{Frequency response curves representing the displacements of the two outer masses in the $n^{\text{th}}$ super cell of a finite acoustic metamaterial (i.e., $u_{b,n}$ and $u_{c,n}$) as well as the displacements of the two internal resonators of the same super cell (i.e., $v_{b,n}$ and $v_{c,n}$), corresponding to: (a) Scenario A, (b) Scenario B, (c) Scenario C, and (d) Scenario D. In these plots, $\omega_c = 1e5$, $n=10$, and shaded regions indicate the different band gaps predicted by the dispersion analysis.}
	\label{fig:FRF_omega_c_100000}
\end{figure}

Figure~\ref{fig:FRF_omega_c_100000} shows the frequency response of the finite metamaterial with the super cell configuration when $m_a=1$, $m_b=m_c=0.3$, $k_a=4.8e9$, $\omega_c=1e5$, and $n=10$. The four plots shown correspond to the four scenarios depicted in Figs.~\ref{fig:Supercell_omega_c_100000_part1} and \ref{fig:Supercell_omega_c_100000_part2} (A, B, C, and D) of the unit cell analysis. In each sub-figure, four different frequency responses are shown which represent the displacements of the two outer masses in the $n^{\text{th}}$ super cell of the finite chain ($u_{b,n}$ and $u_{c,n}$) as well as the displacements of the two resonators of the same super cell ($v_{b,n}$ and $v_{c,n}$). The figures confirm all the various wave propagation patterns observed in the dispersion analysis. Specifically, Fig.~\ref{fig:FRF_omega_c_100000}a shows the presence of an initial low-frequency local resonance band gap followed by a Bragg band gap which houses the second local resonance (i.e., Scenario A). In particular, the frequency response of $v_{c,n}$ shows the vanishing of the local resonance attenuation peak inside the Bragg gap as predicted by the dispersion diagram of the same scenario. Some additional features can be observed such as the truncation resonant peaks within the Bragg band gap and the anti-resonance exhibited by $u_{c,n}$ at the tuning frequency of the second resonator. Truncation resonances are unique to finite periodic systems and cannot be observed in infinite dispersion diagrams \cite{NouhSinghJASA, hussein2015flow}, and they exist in all four transfer functions since they share the same denominator (aka characteristic equation). Anti-resonances are indicative of transfer function zeros which are functions of the sensing location and thus vary from one mass to the next along the chain. Figure~\ref{fig:FRF_omega_c_100000}b shows Scenario B where a merged local resonance band gap is followed by a smaller Bragg band gap. Figure~\ref{fig:FRF_omega_c_100000}c confirms the disappearance of the Bragg gap in Scenario C and that the only band gap in this design is a local resonance band gap (not counting the unbounded stop band at high frequencies). Finally, Figure~\ref{fig:FRF_omega_c_100000}d shows Scenario D which captures the widest merged local resonance band gap with a minuscule Bragg band gap that precedes it. The latter is only captured using a fine resolution in the dispersion diagram is (understandably) hardly noticeable in the frequency response.

\begin{figure}[h!]
\centering
	\includegraphics[width=\textwidth]{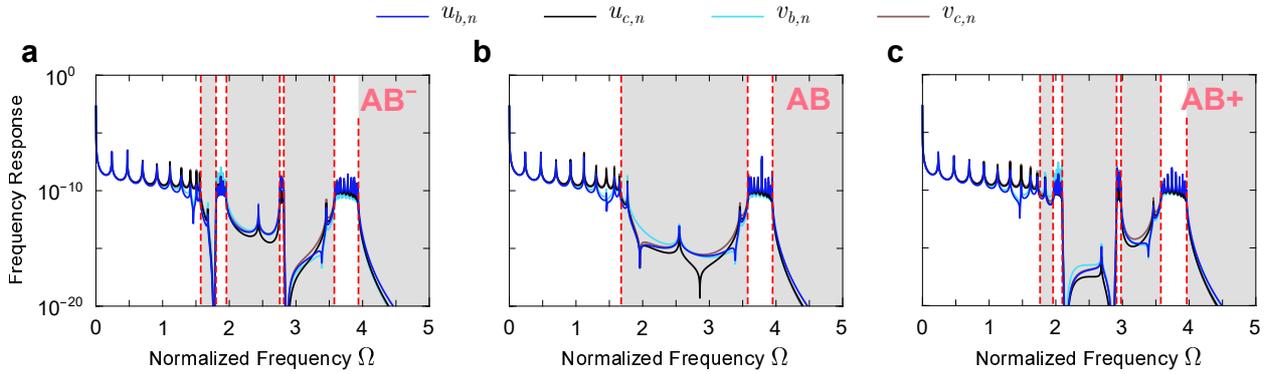}
	\caption{Frequency response curves representing the displacements of the two outer masses in the $n^{\text{th}}$ super cell of a finite acoustic metamaterial (i.e., $u_{b,n}$ and $u_{c,n}$) as well as the displacements of the two internal resonators of the same super cell (i.e., $v_{b,n}$ and $v_{c,n}$), corresponding to: (a) Scenario AB$^-$, (b) Scenario AB, and (c) Scenario AB$^+$. In these plots, $\omega_c = 1.14e5$, $n=10$, and shaded regions indicate the different band gaps predicted by the dispersion analysis.}
	\label{fig:FRF_omega_c_114311_special_case}
\end{figure}

To conclude, Fig.~\ref{fig:FRF_omega_c_114311_special_case} sheds light onto the special scenario AB which corresponds to $\omega_c = 1.14e5$ and $\omega_{b,\text{AB}} = 0.78e4$, and is a result of the intersection of Scenarios A and B as shown in Fig.~\ref{fig:omega_b_over_omega_c}a. The three frequency response plots shown in Fig.~\ref{fig:FRF_omega_c_114311_special_case} correspond to the three dispersion diagrams plotted in Fig.~\ref{fig:omega_b_over_omega_c}b which represent cases AB$^-$, AB, and AB$^+$. These frequency responses show the gradual development of Scenario AB in the finite metamaterial. The behavior perfectly matches the unit cell predictions. As anticipated, the AB$^-$ design at $\omega_b < \omega_{b,\text{AB}}$ (Fig.~\ref{fig:FRF_omega_c_114311_special_case}a) exhibits a low-frequency local resonance band gap corresponding to the first tuning frequency followed by a Bragg band gap and then a high-frequency local resonance band gap corresponding to the second tuning frequency. As soon as $\omega_b = \omega_{b,\text{AB}}$, the two local resonances are embedded within a single wide Bragg band gap, as seen in Fig.~\ref{fig:FRF_omega_c_114311_special_case}b. This scenario is of great interest since the metamaterial exhibits the widest band gap attainable using the super cell configuration. Finally, for the AB$^+$ design at $\omega_b > \omega_{b,\text{AB}}$, the finite metamaterial exhibits a merged local resonance band gap with small and large Bragg band gaps before and after, respectively.

\section{Conclusions}

This paper investigated a range of different functionalities which can be provided by multi-resonator acoustic metamaterials (AMs), both in their infinite and finite manifestations, specifically in the domain of vibration attenuation using frequency band gaps. Both the dispersion analysis and the transfer function approach focused on two main configurations of multi-resonator AMs, namely a unit cell with parallel resonators and a super cell with two distinct resonators which are housed in two masses connected in series. The primary takeaways can be summarized as follows: 

\begin{enumerate}
\item Significant local resonance band gap widening can take place by assembling identical resonators in parallel within each unit cell. Compared to a single-resonator design, both the upper and lower band gap bounds shift as a result of the parallel configuration. Consequently, the band gap expands from both sides, albeit more significantly from the high frequency end. 
     \begin{itemize}
         \item At the unit cell level, the lower and upper band gap bounds, $\Omega_l$ and $\Omega_u$, are both favorably impacted by the addition of the parallel resonators resulting in a downward shift of the acoustic mode and an upward shift of the optical mode. The increase in resonators also adds local resonance modes to the dispersion diagram which are benign and do not interrupt the band gap behavior at the finite level. These modes depict the different relative motions between the individual resonators in the parallel arrangement.
         \item At the finite metamaterial level, a lighter and smaller chain multi-resonator chain can be used to achieve a transmission spectrum which exhibits less resonances in addition to a wider band gap.
         \item Connecting the resonators in series and stacking the in-series chain multiple times in parallel successfully achieves both of the above functions, i.e., the generation of multiple band gaps along the frequency axis in addition to widening each of these gaps. In this hybrid design, the widening effect appears to be more pronounced in the higher frequency gaps than it is in the lower frequency ones, although clearly apparent in all of them.
     \end{itemize}
\item The dual-periodic super cell configuration can be used to generate a wide range of dispersion profiles depending on the choice of both tuning frequencies $\omega_b$ and $\omega_c$. 
    \begin{itemize}
        \item Generally speaking, the super cell exhibits two separate local resonance band gaps corresponding to the two distinct resonators, in addition to a Bragg band gap as a result of its periodic nature. 
        \item The four degrees-of-freedom in the lumped spring-mass model of the super cell yields four dispersion branches and a total of eight solutions to the dispersion relation between $\tilde{k}=\pi$ and $\tilde{k}=0$. Multiple intersections of these solutions for different combinations of $\omega_b$ and $\omega_c$ produce a subset of designs which display unique four dispersion scenarios: A, B, C, and D. The first two scenarios can intersect to generate a special design AB which represents a fifth scenario.
        \item Between the aforementioned five scenarios, a detailed road map is provided which can be used to selectively induce: (1) a smooth (uninterrupted) merge of the two local resonance band gaps with either single or double attenuation peaks, (2) an ultrawide band gap, (3) an unconventional low-frequency Bragg band gap which precedes any other gap along the frequency axis, and finally (4) a complete disappearance of one or more band gap types.
    \end{itemize}
 \end{enumerate}

The presented work paves the way for acoustic metamaterials which rely upon local resonance effects to impact a wider range of practical engineering applications which demand broader vibration attenuation capabilities. While representative designs were shown to emphasize the underlying physics and functionalities that each configuration serves, the framework presented here can be readily implemented in a larger design/parameter space using different optimization methods where the resultant band gaps and their corresponding bandwidth can be tailored to meet specific needs.

\section*{Acknowledgement}

The authors acknowledge the support of this work by the US National Science Foundation through CMMI Award numbers 1904254 and 2021710.


\bibliographystyle{elsarticle-num}
\bibliography{references}

\end{document}